\documentclass[a4paper,11pt]{article}
\usepackage{jcappub}

\usepackage{subfigure}
\usepackage{wrapfig}
\usepackage{xcolor}
\usepackage{hyperref}
\usepackage[normalem]{ulem}

\arxivnumber{2604.26022} 
\title{\boldmath Secondary Dependence of Baryonic Effects on the Density Profile of Dark Matter Halos}

\author[a]{Yikun Wang,}
\author[a]{Idit Zehavi,}
\author[b]{Sergio Contreras,}
\author[c]{Giovanni Aricò,}
\author[d]{Sownak Bose,}
\author[e]{Lars Hernquist}
\affiliation[a]{Department of Physics, Case Western Reserve University, 10900 Euclid Avenue, Cleveland, OH 44106-1715, USA}
\affiliation[b]{Facultad de F\'isica. Universidad de Sevilla. Multidisciplinary Unit for Energy Science, Av. Reina Mercedes s/n 41012 Seville, Spain}
\affiliation[c]{Istituto Nazionale di Fisica Nucleare, Sezione di Bologna, viale Berti-Pichat 6/2, 40127 Bologna, Italy}
\affiliation[d]{Institute for Computational Cosmology, Department of Physics, Durham University, \\South Road, Durham DH1 3LE, UK}
\affiliation[e]{Harvard–Smithsonian Center for Astrophysics, 60 Garden Street, Cambridge, MA 02138, USA}

\emailAdd{idit.zehavi@case.edu}

\abstract{Baryonic physics is anticipated to be a major source of systematic uncertainty in current and future large-scale cosmological surveys. We investigate how baryonic effects on halo density profiles vary with secondary halo properties at fixed halo mass, using the large-volume MillenniumTNG hydrodynamical simulation and its dark matter-only counterpart. We focus on the impact of halo concentration and large-scale environment on the ratio of density profiles of matched halos in the hydrodynamical and dark matter-only simulations. At redshift $\mathrm{z} = 0.0$, we find a strong dependence on halo concentration, especially at lower halo mass ($12.5 < \log(\mathrm{M_h}/h^{-1} \mathrm{M_{\odot}}) < 13.0$), where more concentrated halos exhibit weaker inner enhancement and stronger intermediate-radius suppression at fixed halo mass, with variations reaching $\sim 15\%$ at small scales and decreasing toward larger scales. This trend weakens and reverses at higher halo mass. In contrast, the secondary dependence on large-scale environment is weaker ($\sim 2\%$) and largely scale-independent, with halos in denser regions exhibiting slightly weaker intermediate suppression. By separating internal profile redistribution from total mass suppression, we show that concentration impacts both components, whereas the environmental dependence is primarily associated with an overall mass shift. These secondary dependencies persist at $\rm z = 0.5$ and correlate with variations in internal baryonic properties. We examine additional halo properties, including halo spin and velocity dispersion, and find significant secondary dependence. Overall, our results highlight the important role of secondary halo properties in modulating baryonic effects on halo density profiles, with potential implications for future modeling efforts.}

\begin{document}
\maketitle
\flushbottom

\section{Introduction}
\label{sec:intro}

In the standard $\Lambda$CDM cosmological framework, galaxies reside inside dark matter halos and evolve alongside them (e.g., \cite{White.1996, Wechsler.2018}). The formation and evolution of dark matter halos are primarily governed by gravitational interactions among collisionless dark matter particles, allowing for precise predictions using high-resolution numerical simulations. In contrast, the formation of galaxies is inherently more complex, driven by the interplay between hierarchical structure formation and a variety of intricate baryonic processes that remain difficult to model. The inclusion of collisional and dissipative baryons and the associated feedback mechanisms leads to a significant redistribution of matter in the Universe and introduces substantial complexity to cosmological simulations \cite{Daalen.2011}. This complexity remains a major source of systematic uncertainty for inferences from current and future large-scale weak-lensing surveys (e.g., \cite{Semboloni.2011, Joachim.2015, Huang.2019, Parimbelli.2019}). Consequently, understanding the impact of these baryonic effects is essential both for advancing models of galaxy formation and for robustly constraining cosmological parameters.
 
To better model the galaxy population, a variety of cosmological hydrodynamic simulations have been developed. These simulations describe baryons using either Lagrangian (e.g., \cite{Gnedin.1995, Pen.1995}) or Eulerian methods (e.g., \cite{Cen.1992, Ryu.1993, Teyssier.2002}), and incorporate sub-grid models to implement unresolved baryonic processes, such as star formation, radiative gas cooling, and feedback from active galactic nuclei (AGN) and stellar winds (e.g., \cite{Vogelsberger.2020}). These processes drive intricate interactions and energy exchanges among baryons, thereby regulating the distribution of matter throughout the Universe. Moreover, the gravitational coupling between baryons and dark matter induces a reconfiguration of the dark matter itself, commonly referred to as {\it back-reaction} \cite{Daalen.2011, Gebhardt.2026}. The interplay of these mechanisms produces complex, non-linear modifications to the matter distribution on both internal structures and large scales, deviating significantly from dark matter-only (hereafter DMO) simulations that include only gravitational physics (e.g., \cite{Cui.2012,Cui.2014,Springel.2018,Ferlito.2023, Sorini.2025, Gebhardt.2026}). The magnitude and features of these baryonic effects are highly sensitive to the specific implementations of baryonic physics prescriptions, which vary between simulation projects.

Numerous studies have utilized cosmological hydrodynamical simulations to investigate how baryonic physics modifies the properties of dark matter halos. Several works have shown that baryonic processes generally act to reduce the overall halo masses, as reflected in the suppression of the halo mass function (e.g., \cite{Cui.2012,Cui.2014,Velliscig.2014, Beltz-Mohrmann.2021, Sorini.2022}). Subsequent investigations into correlation functions and density profiles have reported a suppression of the matter distribution by $\sim 20\%$ in the intermediate regions due to feedback, along with significant enhancements in the central regions driven by baryonic condensation and star formation (e.g., \cite{Cui.2014, M.Schaller.2015, Springel.2018}). Additional works focusing on the concentration-mass relation have identified distinct modifications in its shape (e.g., \cite{Beltz-Mohrmann.2021, Shao.2024, Sorini.2025}). Moreover, baryonic feedback has been shown to substantially alter the matter power spectrum, leading to significant suppression on a range of relevant scales (e.g., \cite{Daalen.2011, Chisari.2018, Springel.2018, G.Arico.2020}). 

In parallel, a variety of baryonification models have been developed to emulate the baryonic effects on the cosmic density field using gravity-only cosmological N-body simulations (e.g., \cite{Schneider&Teyssier.2015, Schneider.2019, Schneider.2020, G.Arico.2020, G.Arico.2021, G.Arico.2024, Schneider.2025}). These methods define a parametric mapping from halos in DMO simulations to their modified counterparts according to physically motivated prescriptions, explicitly modeling particle-level distribution of stars, gas, and dark matter components. They provide a flexible and computationally efficient framework for interpreting weak gravitational lensing and other cosmological observables. Alternatively, other works have proposed resummation models to map the baryon fractions of massive halos directly to the ratio of baryonic to DMO matter power spectra, achieving accurate predictions without relying on any free parameters (e.g., \cite{van.2024, van.2025}).

All previous studies have primarily focused on halo mass as the dominant factor governing baryonic effects and as the primary parameter in the baryonification models. However, extensive research into halo and galaxy assembly bias has demonstrated that secondary properties, such as halo concentration, formation time, spin, velocity dispersion and large-scale environment, play a crucial role in shaping the halo clustering and influencing the formation and evolution of galaxies beyond halo mass (e.g., \cite{Gao.2007, Faltenbacher.2010,Mao.2018,  Artale.2018, Zehavi.2018, Zehavi.2019, Bose.2019, Sergio.2019, Xu.2021b, Wang.2025}). For instance, variations in halo concentration and environment have been shown to affect the central galaxy mass and satellite abundance of halos at fixed halo mass \cite{Zehavi.2018, Zehavi.2019, Sergio.2019}, suggesting a corresponding modulation of the total baryonic matter distribution within halos. This naturally raises the question: Do these secondary variations in the galaxy population translate into a measurable dependence in the baryonic effects on halo structure?

Recent studies provide growing evidence supporting this hypothesis (e.g., \cite{Beltz-Mohrmann.2021, Y.Wang.2024, Sunseri.2023, Sims.2026}). Specifically, ref.~\cite{Sunseri.2023} demonstrates that the impact of baryonic feedback varies significantly with cosmic web features, highlighting the importance of modeling baryonic effects across broader environments. Similarly, ref.~\cite{Beltz-Mohrmann.2021} finds that the baryonic modification of the total halo mass depends on the large-scale environment. Ref.~\cite{Elbers.2025} further extends the analysis to the influence of halo concentration. Regarding scale-dependent measurements, ref.~\cite{Y.Wang.2024} employs an environment-dependent wavelet power spectrum to show that the baryon fraction decreases with increasing environmental density, attributed to efficient gas ejection by feedback mechanisms in over-dense regions. However, among these studies, some lack strict control over halo mass, while others lack a detailed analysis of halo density profiles, hindering the isolation of genuine secondary dependencies of baryonic effects on the matter distribution.

In this paper, we explore the secondary dependencies of baryonic effects on halo density profiles using the state-of-the-art MillenniumTNG cosmological hydrodynamic simulations \cite{C.Hernández-Aguayo.2023}. Distinct from previous works, we aim to specifically discern the role of halo concentration and large-scale environment at fixed halo mass. We quantify the baryonic effects by computing the ratio between the density profile of halos in the hydrodynamical run and their DMO counterparts. To control for halo mass, we bin halos according to their mass in the DMO simulation. Within each narrow mass bin, we identify sub-samples of halos at the extremes of the secondary property distribution and compare the averaged baryonic effects. Our primary analysis focuses on redshift $z = 0.0$, with an extension to $z = 0.5$ for relevance to weak-lensing applications. Finally, to facilitate the interpretation of our results, we perform two complementary analyses: we decouple the effects on internal matter redistribution from the overall mass suppression by rescaling the halo profiles, and independently examine how secondary signals correlate with specific internal baryonic components and properties.

This paper is organized as follows. In Section~\ref{sec:methods}, we describe the simulation and the methodology of our analysis. The main results are presented in Section~\ref{sec:secondary}, which examines the secondary dependence on halo concentration (Section~\ref{subsec:con}) and large-scale environment (Section~\ref{subsec:env}) at redshift $\rm z=0.0$, followed by an extension to a higher redshift of $\rm z=0.5$ (Section~\ref{subsec:higher_z}). Section~\ref{subsec:normed} analyzes the mass-scaled signals to separate internal redistribution from overall mass suppression. Section~\ref{sec:phys} discusses the secondary dependence of the different baryonic components (Section~\ref{subsec:baryon_compo}), as well as how the total baryonic effects correlate with internal baryonic properties (Section~\ref{subsec:baryonic_property}).  We summarize our results and conclude in Section~\ref{sec:conclusion}. Appendix~\ref{app:scatter} provides complementary tests on the statistical scatter, and Appendix~\ref{app:spin} presents additional results for halo spin and velocity dispersion.  

\section{Numerical Methods}
\label{sec:methods}

\subsection{The MillenniumTNG Simulation}
\label{subsec:MTNG}

In this work, we analyze galaxy and halo samples from the MillenniumTNG project \footnote{\url{https://www.mtng-project.org/}}, a state-of-the-art suite of large hydrodynamic simulations, incorporating a comprehensive model of galaxy formation and evolution \cite{C.Hernández-Aguayo.2023, R.Pakmor.2023, Barrera.2023, Kannan.2023, Hadzhiyska.2023a, Hadzhiyska.2023b, Bose.2023, Contreras.2023, Delgado.2023, Ferlito.2023}. This suite combines the IllustrisTNG project -- "The Next Generation" Illustris suite of hydrodynamical cosmological simulations of galaxy formation \cite{Pillepich.2017, Springel.2018, Marinacci.2018, Nelson.2019} -- with the large volume reached by the iconic dark matter-only Millennium simulation \cite{Springel.2005}, enabling a simulation framework that unites well-tested physical modeling, unprecedented large volume, and fine resolution. Specifically, we employ the largest hydrodynamical simulation of this suite, which tracks the evolution of $4320^3$ dark matter particles and $4320^3$ gas cells in a periodic box with comoving side length of $L = 500 h^{-1} \mathrm{Mpc} \approx 740 \mathrm{Mpc}$. Hereafter, for brevity, we refer to it as the MTNG simulation. The simulation includes $265$ snapshots between redshift $\rm z = 30.0$ to $\rm z = 0.0$. The dark matter and baryon mass resolutions are $1.12 \times 10^8 h^{-1}\rm M_{\odot}$ and $2 \times 10^7 h^{-1}\mathrm{M_{\odot}} $, respectively. The gravitational softening lengths for the dark matter and stars are set to $\epsilon_{\mathrm{DM, *}} = 3.7 \mathrm{kpc}$. For gas cells, an adaptive softening length is employed, which varies with the cell size down to a minimum of $\epsilon_{\mathrm{gas,min}} = 370 \mathrm{pc}$. In this work, we restrict our analysis to scales larger than $10\,h^{-1}\mathrm{kpc}$, well above the spatial resolution, ensuring that our results are unaffected by gravitational softening. The cosmological parameters assumed in the simulation, $\Omega_M = 0.3089$, $\Omega_b = 0.0486$, $\sigma_8 = 0.8159$, $n_s = 0.9667$, and $h = 0.6774$, are consistent with recent Planck values \cite{Planck.2016}. The initial conditions were set at $\rm z = 63$ using second-order Lagrangian perturbation theory, implemented through an updated version of the \texttt{N-GENIC} code embedded within \texttt{GADGET-4} \cite{Springel.2021}. 

To study the impact of baryons on halo density profiles, we need to match halos in the MTNG hydrodynamical simulation with their counterparts in the corresponding dark matter-only  (hereafter DMO) simulation from the MillenniumTNG suite. This DMO simulation shares the same number of dark matter particles, initial conditions, and cosmology, but only consists of dark matter particles and evolves solely under gravity. Its dark matter resolution is $1.32 \times 10^8 h^{-1} \rm M_{\odot}$. The MTNG and DMO simulations were both performed with the quasi-Lagrangian \texttt{AREPO} code (\cite{Springel.2010, Pakmor.2016, Weinberger.2020}) to follow the coupled dynamics of dark matter particles and gas cells. Halos are identified using the friends-of-friends group finder algorithm \cite{Davis.1985}. Within these groups, the simulations further utilize a novel \texttt{Subfind-HBT} substructure finder code to extract the hierarchical substructure to identify subhalos/galaxies \cite{Springel.2021}.

The MTNG simulation adopts the same baryonic physical model of IllustrisTNG \cite{Weinberger.2017, Pillepich.2018a} using identical parameter choices with a few minor modifications, such as the exclusion of magnetic fields and a simplified treatment of metallicity tracking, which do not have any significant impact on the simulation results for galaxies \cite{R.Pakmor.2023}. We have performed analogous measurements using the IllustrisTNG-300 simulation (TNG300) and found qualitatively similar trends to those presented in this work. However, due to the limited simulation volume of TNG300, the number of selected halos is insufficient to yield robust statistics, resulting in significant noise in the measurements. For this reason, we present only the results from the larger-volume MTNG simulation. In this paper, we focus on the halo sample of the hydrodynamical MTNG simulation and the corresponding DMO counterpart at $\rm z = 0.0$ and $\rm z = 0.5$. We adopt a halo mass cut of $\log (\mathrm{M_h} /h^{-1} \mathrm{M_{\odot}}) > 12.5$, motivated by the fact that group- and cluster-scale halos dominate weak lensing and cosmic shear signals, contribute the most to the total baryonic effects on clustering \cite{G.Arico.2021}, and consequently serve as the primary focus of baryonification models.

\subsection{Measuring the baryonic effects}
\label{subsec:measure}

To analyze the baryonic effects on the halo mass distribution, we compare the density profiles of hydrodynamical halos directly with their DMO counterparts. This pairing is provided by the MTNG halo-matching catalog, which links the subhalo IDs between the hydrodynamical and DMO runs. As described in ref.~\cite{Sorini.2025}, the catalog is constructed by matching each DMO halo to the hydrodynamical counterpart that shares the largest fraction of the same $16$ most gravitationally bound particles, following the method used for the IllustrisTNG simulations (e.g., \cite{Lovell.2018, Bose.2019, Nelson.2019}). In practice, while every DMO halo is assigned one hydrodynamical counterpart, a very small fraction of hydrodynamical halos are linked to multiple DMO counterparts. We therefore exclude these non-unique associations to ensure reliable and bijective matched samples. This cut removes only $\sim 0.2\%$ of halos in the mass range considered, and we have verified that excluding these systems has no impact on our results. We emphasize that the halo masses of matched pairs are not identical due to the impact of baryonic effects (see Section~\ref{sec:intro} and Section~\ref{subsec:normed} for details). Since our analysis focuses on the baryonic modifications relative to the DMO simulation, the halo mass of each matched pair, $\rm M_h$, is defined by its DMO mass. Specifically, $\rm M_h \equiv M_{200c}$, which is the mass enclosed within a sphere whose average over-density is $200$ times the critical density $\rm \rho_c$.

We quantify the baryonic effects on a given halo by the ratio of its total matter density profile in the hydrodynamical run to the dark matter density profile of its corresponding DMO counterpart, $\rho (r)/\rho_{\mathrm{dmo}}(r)$. For a given halo population, the representative signal of baryonic effects is therefore computed as the mean of these individual matched halo-pair ratios. Analogously, the back-reaction effects on dark matter are assessed through the same ratio computed for the dark matter component only, where the DMO profiles are rescaled by a factor of $\rm 1 - \Omega_b/ \Omega_m$ to ensure a consistent comparison with the dark matter profiles in the hydrodynamical run. In practice, we down-sample the simulation data by randomly retaining $3\%$ of all particles and cells for the density profile measurement, which preserves the necessary resolution while substantially reducing the computational cost. This sub-sampling is performed globally and independently for each component (dark matter, gas, and stars). Following this procedure, a representative DMO halo in our analysis ($13.0 <\log (\mathrm{M_h} /h^{-1} \mathrm{M_{\odot}}) < 13.5$) typically retains $\sim 4000$ dark matter particles within the virial radius, whereas individual halo in the lowest narrow mass bin ($ \log (\mathrm{M_h} /h^{-1} \mathrm{M_{\odot}}) \sim 12.5$) average fewer than $1000$ particles. Although this individual particle count is modest, our analysis focuses on the profile ratio, and the final signals are stacked across a large ensemble of halos within each bin, thereby suppressing statistical noise and ensuring robust measurements down to the smallest resolved radii. We tested alternative sampling fractions of $0.5 \%, 1\%$ and $5\%$, finding that the results remain consistent and the resolution of the measured baryonic effects signal has converged for fractions over $1\%$.

To investigate the impact of secondary properties beyond halo mass, we control for halo mass by carrying out all the analyses in fixed bins of halo mass. Selecting an appropriate bin size is critical for obtaining reliable measures of the secondary effects, as overly fine bins increase statistical noise while overly broad bins might leave residual mass dependence. To achieve an optimal balance, we perform measurements in bins of $0.05$ dex and then average the signals over $10$ consecutive non-overlapping bins, representing the results within the $0.5$ dex halo mass interval. This procedure reduces statistical fluctuations associated with the finer binning while reliably measuring the signal for fixed halo masses. We check the robustness of the binning scheme by assessing the residual mass dependence, quantified as the mean ratio of the baryonic effects of the $20\%$ highest- and lowest-mass halos to those of the full sample within each narrow mass bin, averaged over all bins spanning the $0.5$ dex range. Our method yields negligible residual mass dependence. We further test alternative choices, including averaging over $50$ ultra-narrow $0.01$ dex bins, $5$ broader $0.1$ dex bins, and a single broad $0.5$ dex mass bin. The ultra-narrow option introduces significantly higher statistical noise, whereas the broader bins reduce noise but produce stronger residual mass dependence with similar levels of statistical uncertainties.

\begin{figure}[htbp]
\centering
    \includegraphics[width=.6\columnwidth]{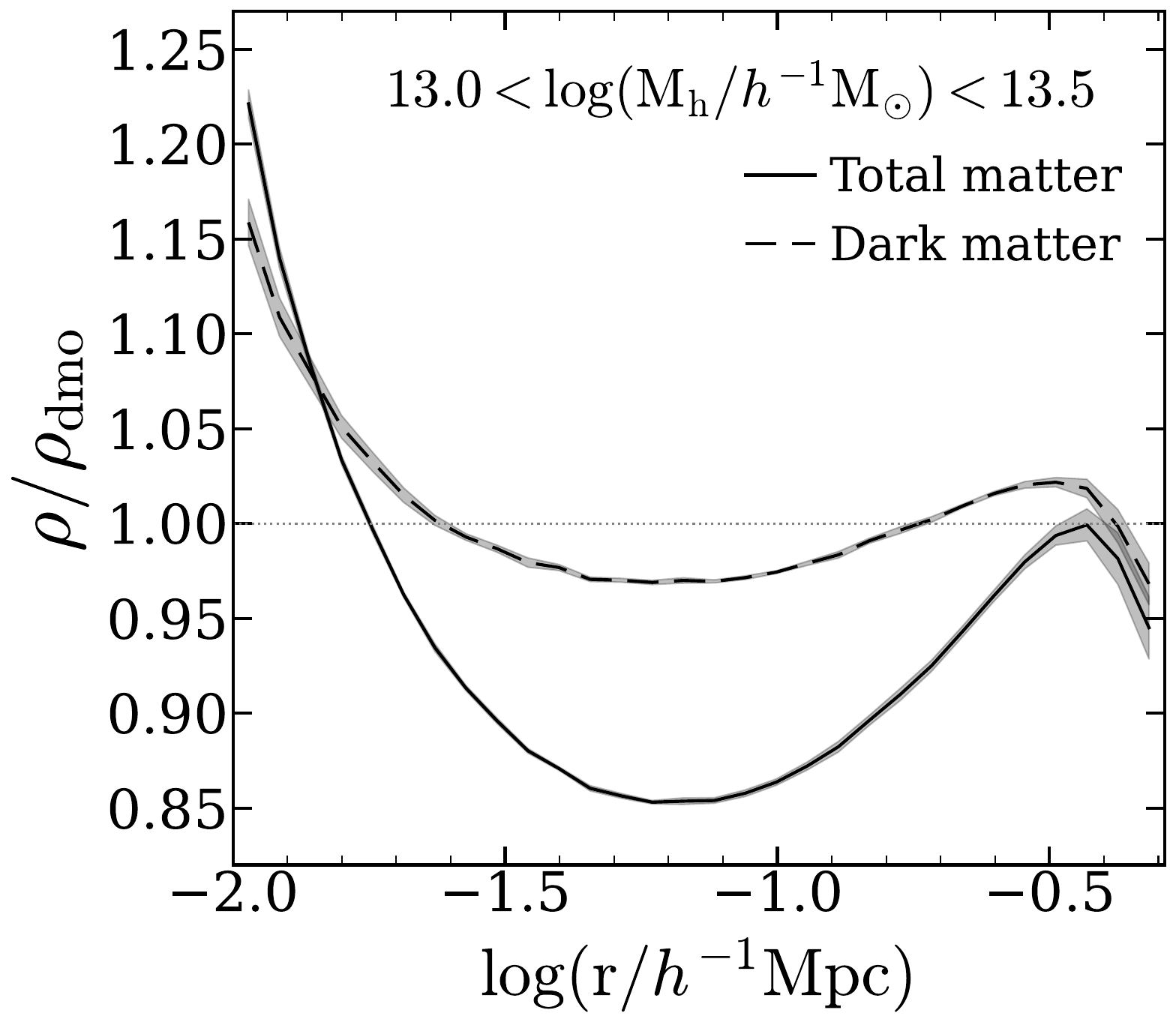}
    \caption{Baryonic effects on halo density profiles for halo mass of $13.0 < \log(\mathrm{M_h} / h^{-1} \mathrm{M_{\odot}})<13.5 $. The solid line displays the average ratio between the density profiles of the total matter in the hydrodynamical halos and their matched DMO counterparts, while the dashed line shows the corresponding ratio for the dark matter component alone. These ratios are computed by averaging the measurements from $10$ halo mass bins of $0.05$ dex width (see Section~\ref{subsec:measure}). The shaded regions represent the statistical uncertainties on the mean of these $10$ measurements. }
    \label{fig:bar}
\end{figure}

Figure~\ref{fig:bar} illustrates the overall baryonic effects on halo density profiles as a function of scale for all halos with mass of $13.0 < \log (\mathrm{M_h} /h^{-1}\mathrm{M_{\odot}}) < 13.5$. The solid line shows the mean ratio between the total matter density profiles of the matched halo pairs, obtained using the procedure described above. The dashed line displays the same measurement for the dark matter components only. The radial range of the profile measurement is truncated at approximately the averaged virial radius, $\rm R_{200 c}$, of the halo in the given mass range, as our analysis focuses solely on the internal halo density profile. Shaded regions indicate the statistical uncertainties on the mean of the ten measurements, which constitute the dominant source of error and also effectively incorporate the uncertainties on the measurements in the narrow mass bins. Although individual halo pairs within each narrow mass bin exhibit considerable scatter, the resulting errors on the mean are significantly smaller due to the large number of halos in each bin. Therefore, the secondary dependence signals explored in this paper do not serve as deterministic predictions for single halos, whereas the population-level trends remain statistically robust (see Appendix \ref{app:scatter} for more details). Accordingly, we report only the uncertainties on the mean of the ten measurements throughout the paper.

As evident in Fig.~\ref{fig:bar}, the halos in the hydrodynamical simulation exhibit, on average, an enhancement of the density profile by over $20 \%$ on small scales of $10 h^{-1}\mathrm{kpc }$, and a suppression by about $15 \%$ on intermediate scales of $\sim 30 - 100 h^{-1}\mathrm{kpc}$, in qualitative agreement with \cite{Springel.2018}. The inner enhancement is primarily driven by the presence of central galaxies, whereas the intermediate suppression mainly arises from the gas ejection by AGN feedback. These mechanisms also induce a back-reaction effect on the dark matter component, as indicated by the dashed line.

In order to assess whether the baryonic effects depend on additional properties at fixed halo mass, we select halo sub-samples based on the secondary property of interest within each $0.05$ dex mass bin. Specifically, in each narrow halo mass bin, we rank the halos by their secondary property and identify those in the top and bottom $20\%$ of the distribution. We note that each selected halo belongs uniquely to a single $20\%$ sub-sample, since the adjacent mass bins are strictly non-overlapping and the selections are performed independently. This procedure enables a direct comparison of the baryonic effects across halos of similar mass but different secondary properties. Unless otherwise specified, these secondary properties of matched pairs are measured from the DMO counterparts to ensure that our selection baseline is unaffected by baryonic physics (with Section~\ref{subsec:baryonic_property} being the sole exception where hydrodynamical properties are explicitly used). We measure the baryonic effects for halos with the $20 \%$ highest and lowest property within each narrow bin by computing the mean $\rho (r)/\rho_{\mathrm{dmo}}(r)$ for each sub-sample. To focus on the impact of secondary properties, we examine the ratio of the baryonic effects of these sub-samples to that of the full sample. Again, we average the ratios from $10$ consecutive narrow mass bins to characterize the overall trend within the broader $0.5$ dex mass range.

\section{The Secondary Dependence of the Baryonic Effects on Halo Density Profiles}  
\label{sec:secondary}

In this section, we present the main results on the dependence of baryonic effects on secondary properties at fixed halo mass, focusing on halo concentration and large-scale environment. Section~\ref{subsec:con} examines how baryonic effects on halo density profiles vary with the halo concentration across different halo mass bins. Similarly, Section~\ref{subsec:env} presents the results regarding the environment dependence, along with its correlation with the concentration dependence. We extend our analysis to a higher redshift of $z=0.5$ in Section~\ref{subsec:higher_z}. Appendix~\ref{app:spin} presents results for two additional secondary properties: halo spin and velocity dispersion.

\subsection{The Secondary Dependence on Concentration}\label{subsec:con}

The halo concentration is defined as the Navarro-Frenk-White (NFW) concentration parameter, $\rm c = R_{\mathrm{200c}}/R_{\mathrm{s}}$, where $\rm R_{\mathrm{200c}}$ is the virial radius of the halo and $\rm R_s$ is the scale radius \cite{Navarro.1997}. In this work, we use the common proxy for the concentration of each DMO halo as the ratio between the maximum circular velocity of the halo, $\rm V_{\mathrm{max}}$, and the virial velocity, $\rm V_{\mathrm{200c}}$ \cite{Bullock.2001,Prada.2012}: $$\rm c = V_{\mathrm{max}}/V_{\mathrm{200c}}$$ where $\rm V_{\mathrm{200c}}$ is the circular velocity at the virial radius. This velocity-based proxy for concentration is particularly useful when the NFW halo profile is difficult to determine or fit, as it can be computed directly from the properties available in the halo catalogs.

\begin{figure*}
\centering
    \includegraphics[width=\textwidth]{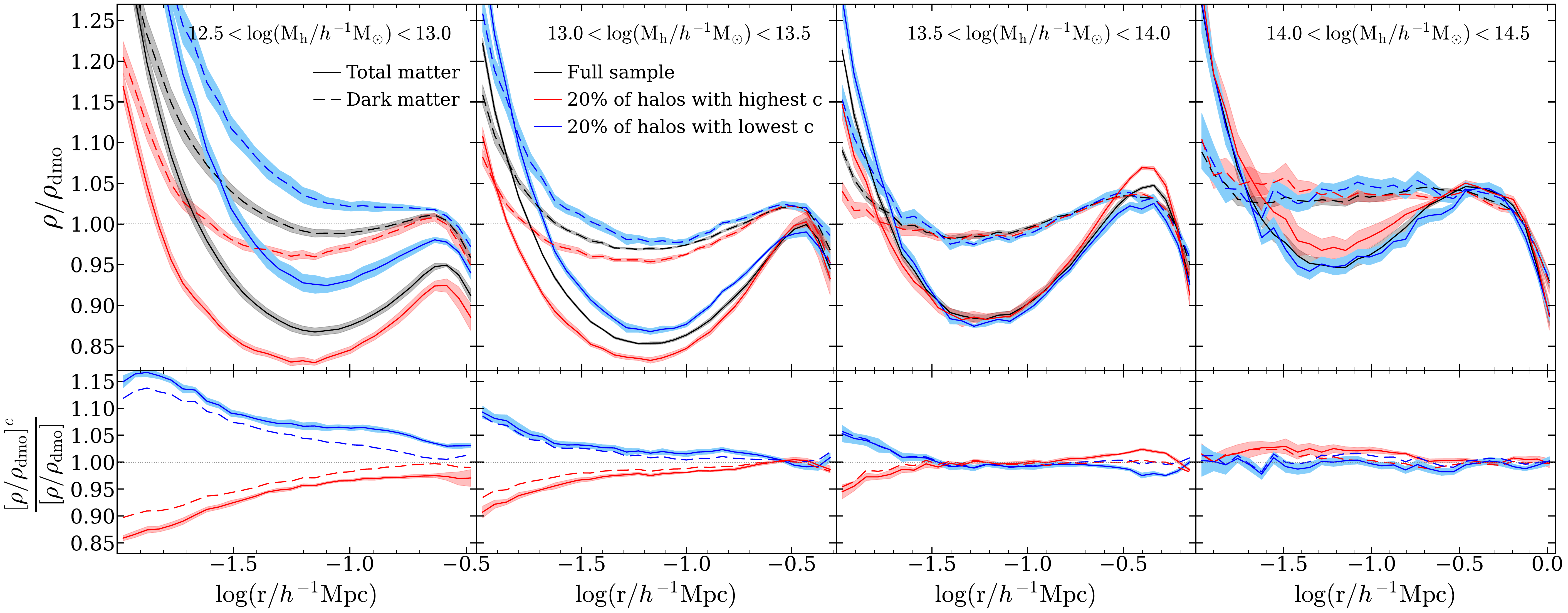}
    \caption{Concentration dependence of baryonic effects on halo density profiles at $\rm z = 0.0$. Each column corresponds to a different halo mass range, as labeled. The top panels show the averaged ratios of matter density profiles, $\rm \rho  (r)/ \rho_{\mathrm{dmo}} (r)$, between each matched halo pair in the hydrodynamical and DMO simulations, with solid and dashed lines indicating the signals for the total matter and the dark matter component, respectively. Black lines denote all matched halo pairs, while red and blue lines represent the $20\%$ most and least concentrated subsets at fixed halo mass, as labeled. The bottom panels display the ratios of the averaged $\rm \rho  (r)/ \rho_{\mathrm{dmo}} (r)$ in these extreme subsets relative to the full sample at fixed halo mass. Again, solid and dashed lines refer to the total matter and dark matter components. In all panels, the upper limit of the $x$-axis corresponds to the mean virial radius, $\rm R_{200c}$, of halos within each mass range. All results are computed by averaging the measurements from $10$ halo mass bins of $0.05$ dex width and shaded regions represent statistical uncertainties on the mean (see Section~\ref{subsec:measure}).}
    \label{fig:c}
\end{figure*}

Figure~\ref{fig:c} shows how the baryonic effects on the density profile of halos vary with the concentration in the MTNG simulation sample for different halo mass bins, at redshift of $\rm z = 0.0$. The top panels show the baryonic effects, comparing the density profiles of matched halos in the hydrodynamical simulation and the corresponding DMO run. The solid lines represent the mean ratios for the total matter, and the dashed lines show those for the dark matter components alone. In each mass interval, we compare the baryonic effects for the full halo sample (black) as well as for the subsets of $20\%$ most concentrated halos (red) and $20\%$ least concentrated halos (blue). We remind the reader that the selection of $20\%$ extreme concentration distribution is performed in $0.05$ dex mass bins. The measurements of baryonic effects for both the full sample and the two concentration-based sub-samples are then averaged across $10$ consecutive narrow bins to represent the result for each broader $0.5$ dex mass interval. The density profiles are computed over the radial range from $10 h^{-1} \mathrm{kpc}$ to the mean virial radius of all halos in the corresponding mass interval. The shaded regions represent the uncertainty on the averaged value from the $10$ measurements. The bottom panels of Fig.~\ref{fig:c} show the ratios of the baryonic effects of the concentration-selected samples to those of the full sample. We note that the red and blue lines in the bottom panels are not derived directly from the top panel, but rather by averaging the ratio of $\rho (r)/ \rho_{\mathrm{dmo}} (r)$ for the $20\%$ most/least concentrated halos and all halos within the $10$ narrower mass bins. Again, the shaded regions show the standard error on the mean value of the $10$ measurements for each sample, as described in Section~\ref{subsec:measure}.

We begin by examining the overall baryonic effects as a function of halo mass, as shown in the top panels of Fig.~\ref{fig:c}. The inner profiles of the MTNG halos are amplified, while the intermediate regions are suppressed (as can also be seen in Fig.~\ref{fig:bar}), and vary somewhat with halo mass. On small scales ($\mathrm{r} \lesssim 20 h^{-1}\mathrm{kpc}$), the baryonic effects significantly enhance the density profiles of both total matter and dark matter by up to $\sim 20\%$, increasingly so toward smaller radii. In the most inner regions, the relative enhancement of the total profile exceeds that of the dark matter component, primarily due to the contribution of central galaxies in the hydrodynamical sample. For larger halo masses, this small-scale enhancement in the total profile weakens only slightly, whereas the dark matter component shows a much stronger dependence on halo mass, leading to a growing discrepancy between the effects on total matter and dark matter profiles. In the intermediate regions on scales of $20$ - $100 h^{-1}\mathrm{kpc}$, the total density profile is suppressed by about $5$-$15\%$ through the four mass ranges. This suppression is mainly driven by AGN feedback, which injects energy and momentum into the surrounding medium and expels gas from halos. The effect diminishes with increasing halo mass, suggesting a weaker impact of the baryonic feedback on more massive halos. This trend arises from the fact that more massive halos possess deeper potential wells, which reduce the overall efficiency of AGN feedback (e.g., \cite{Zinger.2020, Mitchell.2020, Wright.2024}). The corresponding back-reaction on the dark matter components is considerably weaker, showing only mild suppression for the lower-mass halos and an enhancement at the most massive end. At scales near the virial radius, the $\rho/\rho_{\mathrm{dmo}}$ ratio exhibits a mild peak, going above unity for halos with $ \log(\mathrm{M_h} / h^{-1} \mathrm{M_{\odot}}) > 13.5$. This feature may arise from the gas components expelled from the intermediate regions and those infalling from the outskirts \cite{Schneider&Teyssier.2015,Mitchell.2020, Mitchell.2022,Wright.2024}.

We now turn our attention to the baryonic effects in sub-samples selected by halo concentration at fixed halo mass. As shown by the red and blue lines in Fig.~\ref{fig:c}, halos of different concentrations display clear discrepancies, with the trend changing with halo mass. At the low mass end, high-concentration halos generally exhibit significantly smaller hydrodynamical density profiles relative to their DMO counterparts, compared to the low-concentration ones, across most scales. This trend weakens and eventually reverses with increasing halo mass. In the lowest-mass bin ($12.5< \log(\mathrm{M_h} / h^{-1} \mathrm{M_{\odot}}) <13.0$), the $20\%$ least concentrated halos exhibit an enhancement of $\rho /\rho_{\mathrm{dmo}}$ up to $\sim 15\%$ relative to the mean profile of all halos on small scales, whereas the most concentrated halos display a suppression of slightly lower magnitude. At a scale of $\sim 0.1 h^{-1} \mathrm{Mpc}$, the signal decreases to about $8\%$, which remains significant. This indicates that high-concentration halos experience weaker enhancement on their central density profiles and more pronounced suppression in the intermediate regions. For halos in the range $13.0 < \log (\mathrm{M_h} / h^{-1}\mathrm{M_{\odot}}) <13.5 $, the strength of the variation with concentration decreases but remains notable, reaching $\sim 10 \%$  around $0.01 h^{-1} \mathrm{Mpc}$. At higher masses, the trend starts to reverse: for $13.5 < \log (\mathrm{M_h} / h^{-1}\mathrm{M_{\odot}}) <14.0 $ the reversal appears at scales larger than $0.1 h^{-1} \mathrm{Mpc}$, while in the most massive bin ($14.0 < \log (\mathrm{M_h} / h^{-1}\mathrm{M_{\odot}}) <14.5 $), the reversal scale shifts towards the innermost regions. This trend is reminiscent of the reversal in the concentration dependence of halo clustering with increasing halo mass \cite{Wechsler.2006}. In all cases, the corresponding signal for the back-reaction effects on the dark matter profile follows the same qualitative trends, with a slightly lower amplitude, particularly at the low mass end. 

We note that the uncertainties on the secondary dependence signal are relatively large in the most massive bin, primarily due to the limited number of halos available for averaging. The lowest-mass bin ($12.5 < \log (\mathrm{M_h}/h^{-1}\mathrm{M_\odot}) < 13.0$) also shows larger uncertainties compared to other median bins, reflecting greater intrinsic variation in secondary properties toward the low-mass end (see Appendix~\ref{app:scatter}). Additionally, we have found qualitatively similar trends but with much larger noise when using the TNG300-1 simulation and its DMO counterpart. Due to the limited volume of TNG300, the number of halos is insufficient to support robust statistics; consequently, we do not include those results here.

The overall behavior of the concentration dependence is qualitatively consistent with ref.~\cite{Elbers.2025}, which finds that high-concentration halos are suppressed in mass at the low-mass end, whereas the trend reverses toward higher masses. However, ref.~\cite{Elbers.2025} focuses on the total mass ratio between matched halo pairs, whereas our analysis investigates the profile modification. As such, a direct quantitative comparison with previous studies is not straightforward. The observed concentration dependence may arise from the interplay between the depth of the potential wells and the strength of stellar and AGN feedback for halos across different halo mass regimes. We discuss this further in Section~\ref{sec:phys}.

\subsection{The Secondary Dependence on Environment }\label{subsec:env}

We define the large-scale environment of the halos as the dark matter particle density fluctuation field within a sphere of radius $\mathrm{r} = 8 h^{-1}\text{Mpc}$ centered on each halo's central galaxy. To reduce computational cost while maintaining accuracy, we randomly select $0.5 \%$ of the total dark matter particles as the tracer. The environmental density contrast, $\delta$, is then calculated using the standard definition $$\rm \delta = \frac{N - \bar N}{\bar N}$$ where $\rm N$ is the number of tracer particles within $8 h^{-1}\text{Mpc}$ of the halo's (central galaxy) location, and $\rm \bar{N}$ is the mean count across the whole sample. $\rm N$ is measured using \texttt{Corrfunc} \cite{Sinha&Garrison.2020} to count halo–particle pairs. 

The choice of $8 h^{-1}\text{Mpc}$ ensures a smoothing scale much larger than typical halo sizes, thereby probing the surrounding large-scale environment while still capturing a wide range of environments at fixed halo mass. This definition has been widely adopted in galaxy environment studies (e.g., \cite{Croton.2005}), and alternative approaches have also been explored in the literature (e.g., \cite{Muldrew.2012}). We test other smoothing scales, including a top-hat smoothing of $3 h^{-1}\text{Mpc}$ and $5 h^{-1}\text{Mpc}$, as well as a $1.25 h^{-1}\text{Mpc}$ Gaussian smoothing (as adopted in \cite{Xu.2021b}). The top-hat variants yield qualitatively similar results, whereas the Gaussian filter is too small for our purpose and might be dominated by particles inside halos rather than their surrounding environment. In general, our results are not sensitive to the specific working assumptions made here, since the environment measure is used solely to rank halos by environment and to select sub-samples at the extremes.

\begin{figure*}
\centering
    \includegraphics[width=\textwidth]{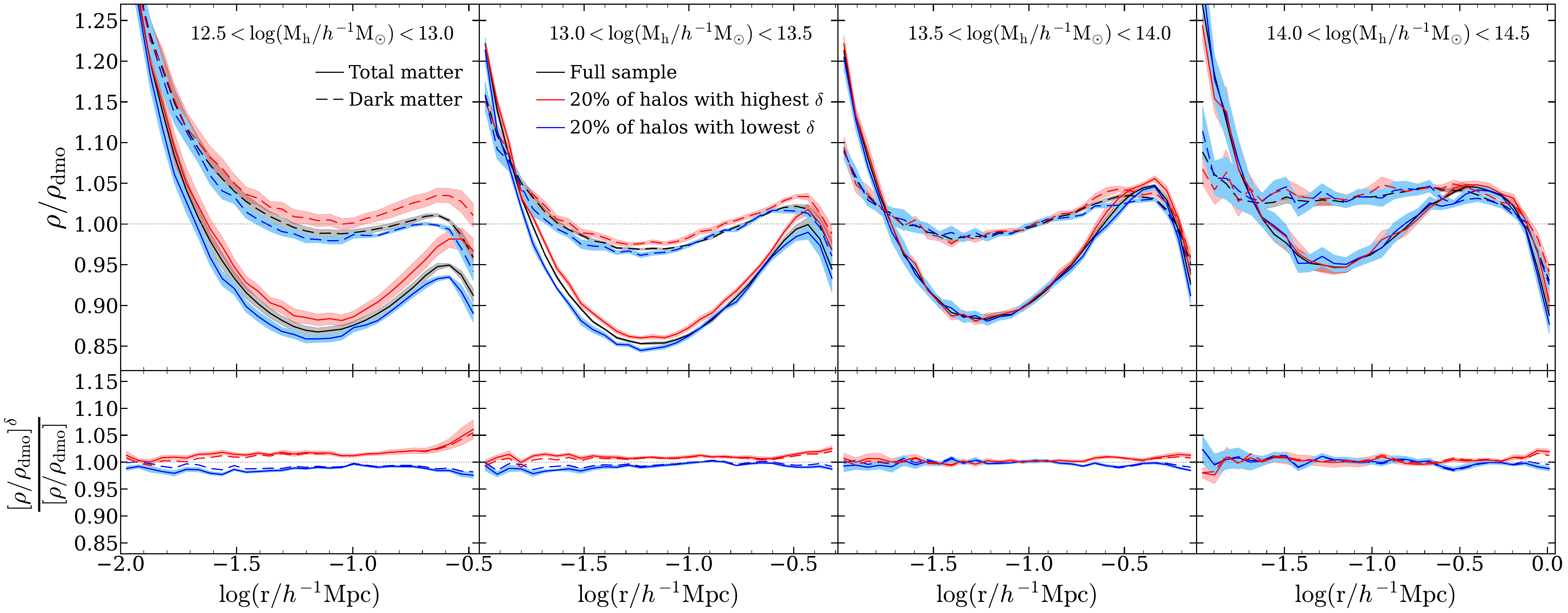}
    \caption{Environmental dependence of baryonic effects on halo density profiles at $\rm z = 0.0$. The same as in Fig.~\ref{fig:c} but for halo samples selected by the environment density at fixed halo mass. The measurements for halo pairs in the $20\%$ densest environments are shown in red, and for the $20\%$ least dense environments are shown in blue. }
    \label{fig:env_0}
\end{figure*}

Figure~\ref{fig:env_0} presents the variation of baryonic effects on the halo density profiles with environment for different halo mass ranges at redshift $\rm z=0.0$. Similar to Fig.~\ref{fig:c}, the top panels show the baryonic effects for both the full halo sample and the $20\%$ subsets residing in the most and least dense environments, within fixed halo mass bins. The bottom panels display the secondary environmental dependence by measuring the ratio between the average baryonic effects of the environment-selected subsets and those of the full sample at fixed mass. All signals are obtained from $10$ measurements in finer halo mass bins of $0.05$ dex width, with the shaded regions representing the statistical uncertainties on the mean value (see Section~\ref{subsec:measure}).

As indicated in Fig.~\ref{fig:env_0}, halos in dense environments exhibit higher $\rho /\rho_{\mathrm{dmo}}$ ratios than those in under-dense regions at the low-mass end. This implies that baryonic processes more effectively enhance central density profiles of halos in the over-dense environments, while the baryonic suppression at intermediate radii is reduced. This effect may be partially attributed to variations in the stellar content of halos at fixed halo mass. In particular, halos in denser environments preferentially host slightly more massive central galaxies and richer satellite populations compared to those in lower-density regions, as indicated by the galaxy assembly bias \cite{Zehavi.2018}. However, the environmental dependence weakens with increasing halo mass and is no longer detectable for $\log (\mathrm{M_h} /h^{-1}\mathrm{M_{\odot}}) > 13.5$. Specifically, for halos with $12.5 < \log (\mathrm{M_h} /h^{-1}\mathrm{M_{\odot}}) < 13.0$, we find a nearly flat secondary dependence signal of about $1.5\%$, and halos with $13.0 < \log (\mathrm{M_h}/h^{-1}\mathrm{M_\odot}) < 13.5$ show a weaker but nonzero signal of similar shape. Toward the center, the signal diminishes slightly, while it strengthens near the virial radius. Although weak, this secondary environmental dependence is still notable relative to the measurement uncertainties. These results further confirm that halo mass is the dominant factor governing the impact of baryonic processes on halos in different environments. The significant environmental dependence of the baryonic effects on matter clustering reported in ref.~\cite{Y.Wang.2024} may largely arise from the strong correlation between halo mass and environment, rather than an independent environmental effect, since their analysis does not fix halo mass.
 
\begin{figure}
\centering
    \includegraphics[width=.8\textwidth]{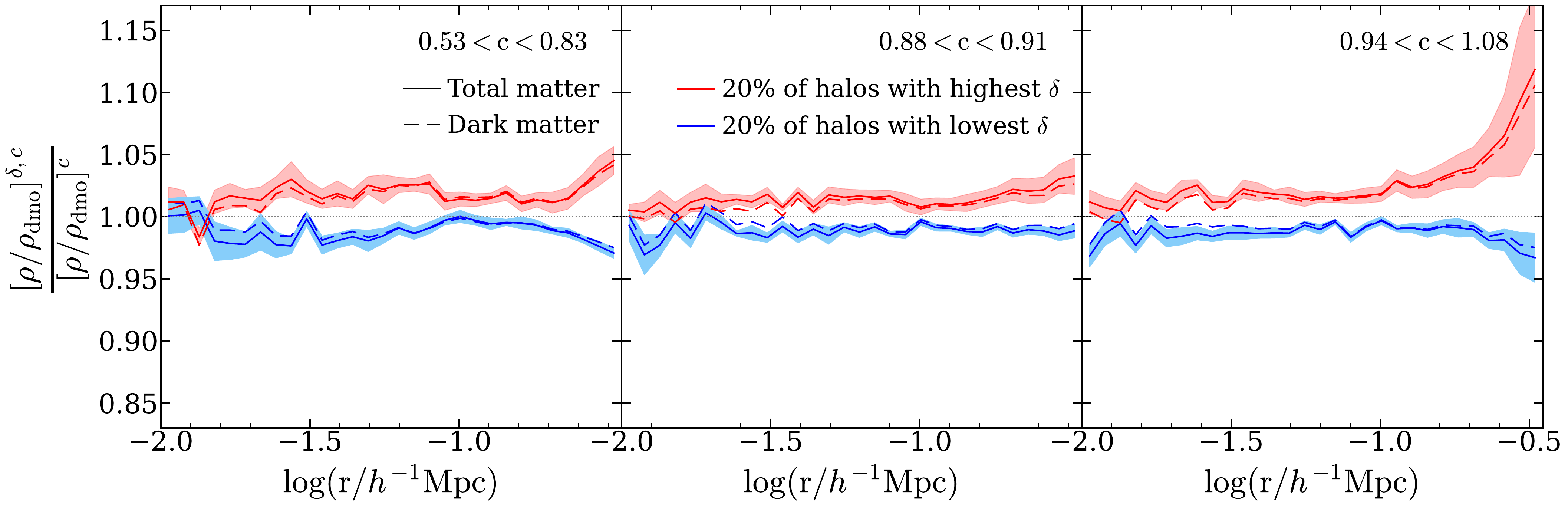}
    \caption{The environmental dependence of baryonic effects on halo density profiles at fixed halo mass and concentration for halos with $12.5< \log(\mathrm{M_h} / h^{-1} \mathrm{M_{\odot}}) < 13.0$. Each column corresponds to a representative concentration bin, which contains approximately $20\%$ of the halos in each $0.05$ dex mass bin. The solid lines represent the ratios of the averaged $\rm \rho  (r)/ \rho_{\mathrm{dmo}} (r)$ of halos in the 20\% most (red) and least (blue) dense environment, relative to that of the full sample within the same concentration bin. Dashed lines correspond to the dark matter components. Shaded regions indicate statistical uncertainties, as in Fig.~\ref{fig:c}.}
    \label{fig:double_env}
\end{figure}

Considering the strong concentration-dependent signal, we further examine whether the secondary environmental dependence of baryonic effects is correlated with concentration. Figure~\ref{fig:double_env} shows the ratio of $\rho /\rho_{\mathrm{dmo}}$ for the $20\%$ halos in dense/under-dense regions relative to all halos with $12.5 < \log (\mathrm{M_h} / h^{-1}\mathrm{M_{\odot}}) <13.0 $ for representative bins of concentration. When fixing concentration, the environment dependence persists with similar trends as seen in Fig.~\ref{fig:env_0}, which demonstrates that the environment dependence is largely independent and not driven by the concentration dependence. This may seem counterintuitive, given that more concentrated halos preferentially reside in denser environments (e.g., \cite{Avila.2005, Wechsler.2006}). However, the correlation between the concentration and environment is negligible in the mass range of our analysis, and the environmental dependence of concentration is weak compared to the intrinsic scatter in concentration at fixed halo mass \cite{Maccio.2007}.

\begin{figure}
\centering
    \includegraphics[width=.6\textwidth]{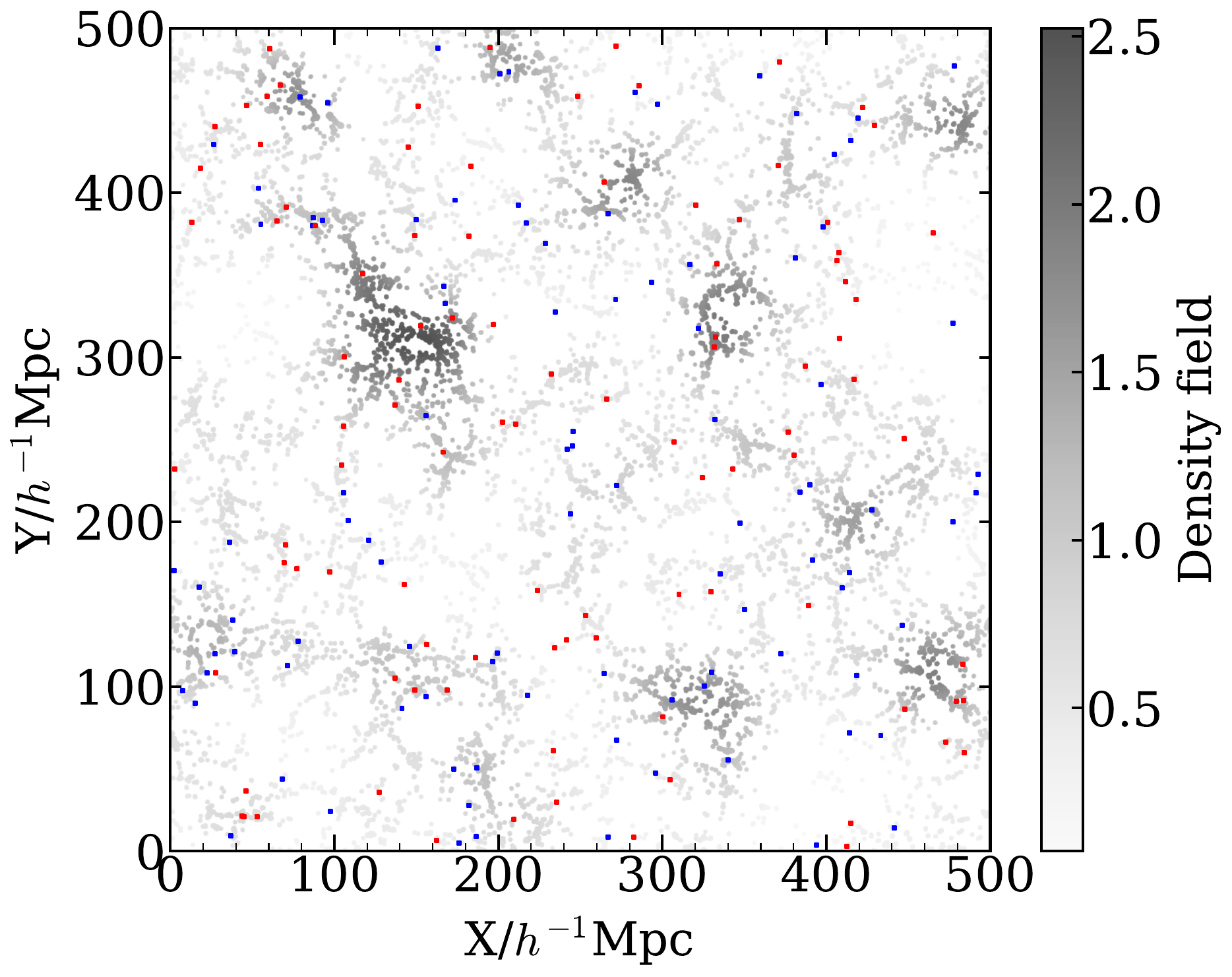}
    \caption{A $500 h^{-1}\mathrm{Mpc} \times  500 h^{-1}\mathrm{Mpc} \times 20 h^{-1} \mathrm{Mpc}$ slice from the DMO simulation showing the spatial distribution of halos that exhibit extreme relative profile ratio, $\rm \langle[\rho(r)/\rho_{dmo}(r)]/ [\rho(r)/\rho_{dmo}(r)]_{mean}\rangle_r$, in their hydrodynamical counterparts, within the halo mass bin $12.50< \log(\mathrm{M_h}/h^{-1} \mathrm{M_{\odot}}) < 12.55$. Note that the red and blue dots denote halos with $20\%$ highest and lowest relative profile ratio, respectively. We emphasize that this classification is distinct from the high/low secondary property classifications represented by the similar color schemes in other figures of this paper. The background grey dots present the underlying dark matter density field. }
    \label{fig:slice}
\end{figure}

Although the secondary dependence on environment is weak in our analysis, its imprint on the spatial distribution of halos with varying baryonic effects can still be visually discerned. Analogous to the ratio in the bottom panel of Fig.~\ref{fig:env_0}, we adopt a relative profile ratio as a proxy for the overall deviation in the baryonic modification of individual halos. Specifically, it is defined for each halo pair as the ratio of its $\rho(r)/\rho_{\mathrm{dmo}}(r)$ to the sample mean, averaged over the radial range from $10^{-2} h^{-1}\mathrm{Mpc}$ to $\rm R_{200c}$. A higher relative profile ratio indicates an overall upward shift relative to the mean profile ratio, which can arise from a stronger central condensation, a weaker intermediate suppression, or a combination of both. Figure~\ref{fig:slice} illustrates the spatial distribution of DMO halos with extreme relative profile ratio in the halo mass bin $12.50< \log(\mathrm{M_h}/h^{-1} \mathrm{M_{\odot}}) < 12.55$, in a slice of the simulation. The background illustrates the underlying dark matter density field. Halos in the top 20\% of this distribution are shown as red dots, while those in the bottom 20\% are shown in blue. Comparing the two populations, we find that denser regions, particularly near clusters, host a slightly larger fraction of halos with higher overall matter retention due to baryonic effects. The difference, however, remains subtle, consistent with the small amplitude of the measured secondary environmental dependence.

Drawing an analogy to galaxy assembly bias studies, we caution the reader that the small secondary dependence on environment observed in our analysis should not necessarily be considered negligible. As found in previous studies of assembly bias, the large-scale environment is known to have a much weaker influence on halo occupancy variation compared to halo concentration, but nonetheless plays a dominant role in driving galaxy assembly bias in the large-scale galaxy clustering \cite{Zehavi.2018, Xu.2021b}. Consequently, even a weak secondary environmental dependence of baryonic effects at the individual halo level may translate into a measurable impact on large-scale clustering. Supporting this view, ref.~\cite{Beltz-Mohrmann.2021} also demonstrates that applying an environment-dependent halo mass correction to DMO halos significantly improves the recovery of large-scale halo clustering of the hydrodynamical sample. With more detailed analysis, our results could similarly inform the inclusion of environmental dependence in more refined baryonification models in the future.

\subsection{Higher Redshift} \label{subsec:higher_z}

Up to this point, our analysis has been limited to examining the secondary dependence of the baryonic effects at redshift $\rm z = 0.0$. Previous studies of the baryonic correction model have shown a clear redshift dependence of baryonic effects (e.g., \cite{Chisari.2018, G.Arico.2020,G.Arico.2021}). In particular, \cite{G.Arico.2020} demonstrated that the baryonic parameters evolve with redshift when fitting the clustering, with the characteristic host halo mass shifting due to the evolution of the halo mass function. Motivated by those works, we extend our analysis to $\rm z = 0.5$, a redshift representative of galaxy lensing surveys, particularly for galaxy-galaxy lensing and cosmic shear studies. Most current cosmological surveys (e.g., \cite{KIDS, Heydenreich.2025, EUCLID.2025}) target galaxies within $0.0 < z < 1.0$, making $\rm z = 0.5$ a natural choice.

\begin{figure*}
\centering
    \includegraphics[width=\textwidth]{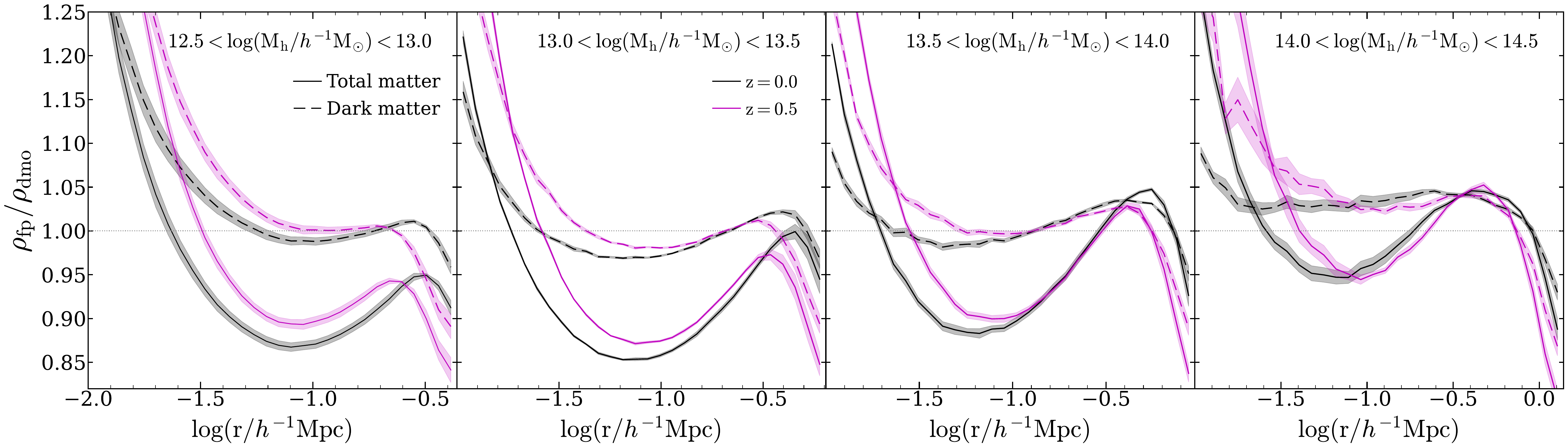}
    \caption{Baryonic effects on the halo density profiles at redshift $\rm z = 0.5$ (purple), compared with $\rm z = 0.0$ (black). Each panel shows a different mass bin, as labeled. The solid lines show the average $\rm \rho  (r)/ \rho_{\mathrm{dmo}} (r)$ for the total matter of all matched halo pairs at fixed halo mass, while the dashed lines correspond to the dark matter component. Other features are the same as in Fig.~\ref{fig:bar}.}
    \label{fig:bar_0.5}
\end{figure*}

Figure~\ref{fig:bar_0.5} compares the impact of baryons on the halo density profile, at $\rm z = 0.5$ and $z= 0.0$, as shown in the magenta and black lines, respectively. Again, solid lines denote the total density profile, while dashed lines show the dark matter component, and the shaded areas indicate the statistical uncertainties. We find that both the shape and amplitude of the signals evolve with redshift. In the inner regions where the baryonic components enhance the density profile, the $\rho /\rho_{\mathrm{dmo}}$ signal at $\rm z=0.5$ is shifted to larger scales by about $0.15$-$0.2$ dex relative to $\rm z = 0.0$ across all mass bins. At intermediate scales, the suppression of the density profile is somewhat weaker at $\rm z = 0.5$ for halos with $\log (\mathrm{M_h} / h^{-1}\mathrm{M_{\odot}}) <14.0 $, but the overall trend remains similar. This redshift evolution may be understood in terms of the concentration dependence of baryonic effects established above: at fixed halo mass, lower-concentration halos exhibit stronger baryon-induced inner-density enhancement and weaker intermediate suppression. Since halos at higher redshift tend to be less concentrated for the same halo mass, the trends seen at $\rm z = 0.5$ are qualitatively consistent with this concentration-driven behavior. Nevertheless, it may also reflect the redshift evolution in the baryonic contents, feedback history, and other evolving halo and galaxy properties.

\begin{figure*}
\centering
    \includegraphics[width=\textwidth]{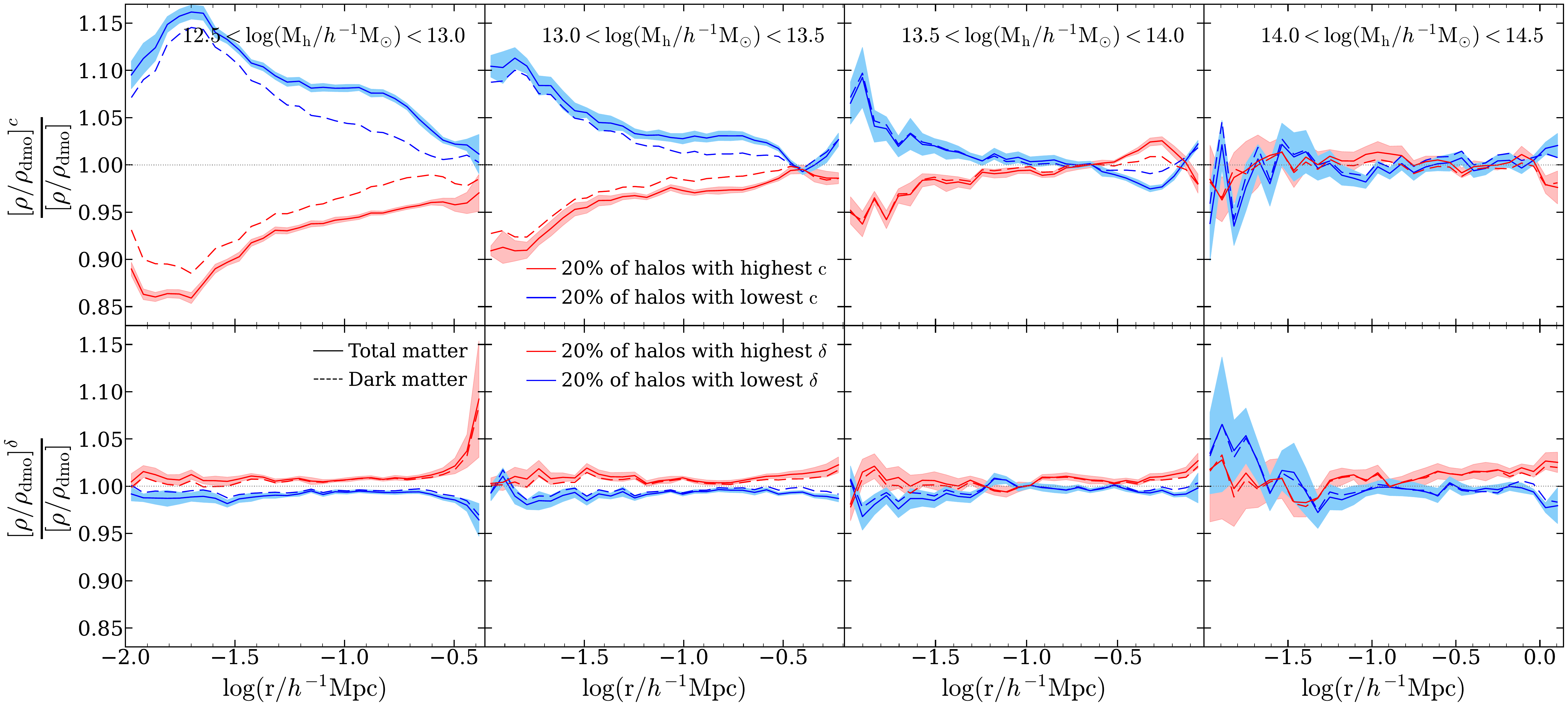}
    \caption{The secondary dependence of baryonic effects on halos at $\rm z = 0.5$, for concentration (top) and environment (bottom). All features are the same as the bottom panels of Fig.~\ref{fig:c} and Fig.~\ref{fig:env_0}. }
    \label{fig:all_0.5}
\end{figure*}

Figure~\ref{fig:all_0.5} shows the secondary dependence of the baryonic effects at $\rm z = 0.5$, where we extract the signal by taking the ratios of $\rho /\rho_{\mathrm{dmo}}$ in sub-samples selected by secondary properties (concentration and environment) to that of the full sample at fixed mass. The overall signals closely resemble those measured at redshift zero. The trends with concentration at $\rm z = 0.5$ (top panels) remain qualitatively similar to those at $\rm z=0.0$ (Fig.~\ref{fig:c}) with a small variation. In the two least massive bins, the peak in the inner regions shifts slightly to larger scales, consistent with the overall evolution of the baryonic effects. Despite the overall reduction of feedback at the higher redshift, the concentration dependence of baryonic effects remains equally significant at $z = 0.5$. For the two massive bins, the reversal of the trend also occurs at larger scales than at $z=0.0$. As for the environmental dependence, the bottom panels of Fig.~\ref{fig:all_0.5} present slightly weaker signals for halo masses between $12.5 < \log (\mathrm{M_h} / h^{-1}\mathrm{M_{\odot}}) <13.5 $ and somewhat stronger but noisier signals at higher masses, compared to those at $z=0.0$ (Fig.~\ref{fig:env_0}).

In summary, the secondary dependence signals remain significant at higher redshift, showing only mild evolution in amplitude and scale. This further emphasizes the importance of accounting for the influence of secondary properties in the baryonic modification over the redshift interval $\rm 0.0<z<1.0$. For all of our results, we caution that the absolute amplitudes of the signals reported here may depend on the specific physics prescription of the hydrodynamical simulations. The strength of baryonic effects is known to vary considerably across different models (e.g., \cite{Beltz-Mohrmann.2021, G.Arico.2021, Y.Wang.2024}). For this reason, our focus is on the global trends revealed by the analysis, rather than the precise numerical values.

\section{Secondary Dependence of the Baryonic Effects on Mass-scaled Halo Profiles} \label{subsec:normed} 

Previous studies have found that baryonic effects generally act to reduce the total halo mass in hydrodynamic simulations relative to their DMO counterparts, with the level of suppression depending primarily on halo mass (e.g., \cite{Cui.2012, Cui.2014, Castro.2021}). More recent work has begun to explore how this overall mass suppression varies with secondary halo properties. For instance, \cite{Beltz-Mohrmann.2021} find that halos in high-density environments are slightly more massive relative to their DMO counterparts than those in low-density environments. Ref.~\cite{Elbers.2025} further explores the impact of halo concentration, reporting that at lower halo masses, more concentrated halos experience stronger mass suppression, whereas this trend reverses at higher halo masses. Motivated by these results, we aim to disentangle the impact of baryons on the internal mass distribution from their effect on the overall mass shift. This allows us to better isolate and interpret the scale-dependent secondary dependencies of baryonic effects.

\begin{figure*}
\centering
    \includegraphics[width=\textwidth]{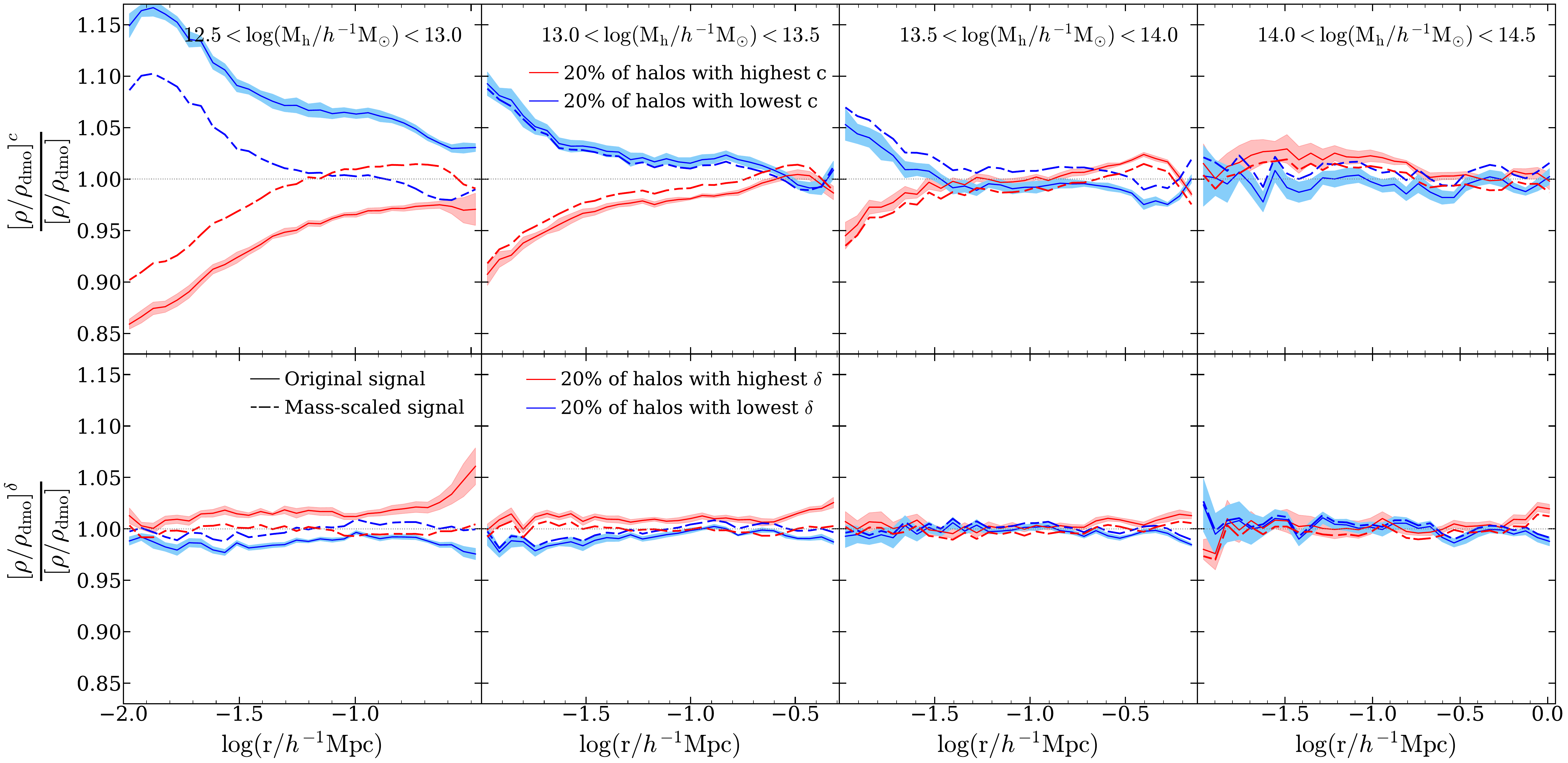}
    \caption{Secondary dependence of baryonic effects on the halo profiles scaled by the halo mass ratio $ \mathrm{M_h}/\mathrm{M_{h, dmo}}$ at $\rm z = 0.0$. Each column corresponds to a different halo mass bin. The dashed lines represent the ratio of the average mass-scaled profile, $(\rho / \rho_{\mathrm{dmo}}) / (\mathrm{M_h}/\mathrm{M_{h, dmo}})$, for the 20\% of halos with the highest (red) and lowest (blue) concentration (top) and environment density (bottom), relative to the full halo population at fixed mass. The solid lines show the original secondary dependence signals for comparison. All the results are averaged across $10$ bins of $0.05$ dex in halo mass (see Section~\ref{subsec:measure}). Shaded regions indicate the statistical uncertainties on the mean value of the original signals.}
    \label{fig:norm}
\end{figure*}

In this section, we rescale the density profile of each hydrodynamical halo by dividing it by the mass ratio $\rm M_{h}/ M_{h, dmo}$ of the corresponding matched halo pair. In this way, the mass of each hydrodynamical halo is effectively normalized to match that of its DMO counterpart. The resulting scaled density profile ratio exhibits the internal mass redistribution induced by baryonic effects, while excluding the contribution from the overall mass shift. Figure~\ref{fig:norm} presents the mass-scaled signals associated with halo concentration and the large-scale environment, shown as dashed lines, alongside the original signals in solid lines. The shaded regions indicate the statistical uncertainties of the original measurements. Since the uncertainties of the mass-scaled signals are comparable, their error bars are omitted for clarity.

As shown in the top panels of Fig.~\ref{fig:norm}, the concentration dependence remains significant after the mass-scaling, though with a somewhat reduced amplitude. Across the three mass bins spanning $12.5< \log(\mathrm{M_h} / h^{-1} \mathrm{M_{\odot}}) < 14.0$, the mass-scaled signals exhibit similar amplitudes, with only slight shifts in shape. This suggests that the variation in the original secondary signal amplitude across halo mass is primarily driven by the concentration dependence of baryonic effects on the total mass shift. Specifically, for halos with $ \log(\mathrm{M_h} / h^{-1} \mathrm{M_{\odot}}) < 13.5$, it can be inferred that more concentrated halos tend to undergo stronger total mass suppression due to baryonic effects. This trend weakens with increasing halo mass, and reverses in the $13.5< \log(\mathrm{M_h} / h^{-1} \mathrm{M_{\odot}}) < 14.0$ range, where more concentrated halos exhibit a slight mass enhancement relative to the average. For the most massive bin, the mass-scaled signal is nearly absent, reflecting a negligible influence of concentration on the matter redistribution from baryonic effects. In contrast, the signal associated with the environmental dependence nearly vanishes after the mass scaling, as can be seen in the bottom panels of Fig.~\ref{fig:norm}. This indicates that the observed secondary environmental dependence primarily reflects an overall mass shift in $\rm M_{h}$, with halos in lower-density regions experiencing slightly stronger mass suppression than those in denser environments.

Overall, the secondary dependencies of the total halo mass shift inferred from our mass-scaling analysis are qualitatively consistent with previous findings in ref.~\cite{Beltz-Mohrmann.2021} and ref.~\cite{Elbers.2025}. We emphasize that the mass-scaled signals do not alter any of our primary results and conclusions, as the overall mass shift is an integral component of the baryonic effects under investigation. However, this supplementary analysis provides insight into the nature of the secondary dependence. Specifically, the environmental dependence is found to manifest almost entirely through an overall normalization of the density profile, with minimal scale dependence, suggesting that it may be effectively modeled through a simple mass-normalization prescription in baryonification models. In contrast, the dependence on concentration is more complex, arising from two contributing effects: (1) a mass-independent restructuring of the density profile, and (2) a scale-independent overall mass shift that varies with halo mass. These distinctions help disentangle the complex signals of secondary dependencies and may facilitate the development of more accurate and efficient baryonification models.

\section{Correlation with Baryonic Properties} \label{sec:phys}

We have shown in the previous sections how the overall baryonic effects on halo density profiles depend on halo concentration and environment, at fixed halo mass. To gain a deeper insight into the underlying physics behind these signals, we further investigate the secondary dependence of the individual baryonic component in Section~\ref{subsec:baryon_compo}, and examine how baryonic properties modulate the total density profile modification in Section~\ref{subsec:baryonic_property}.

\subsection{The Secondary Dependence of the Baryonic Components} \label{subsec:baryon_compo}

\begin{figure*}
\centering
    \includegraphics[width=\textwidth]{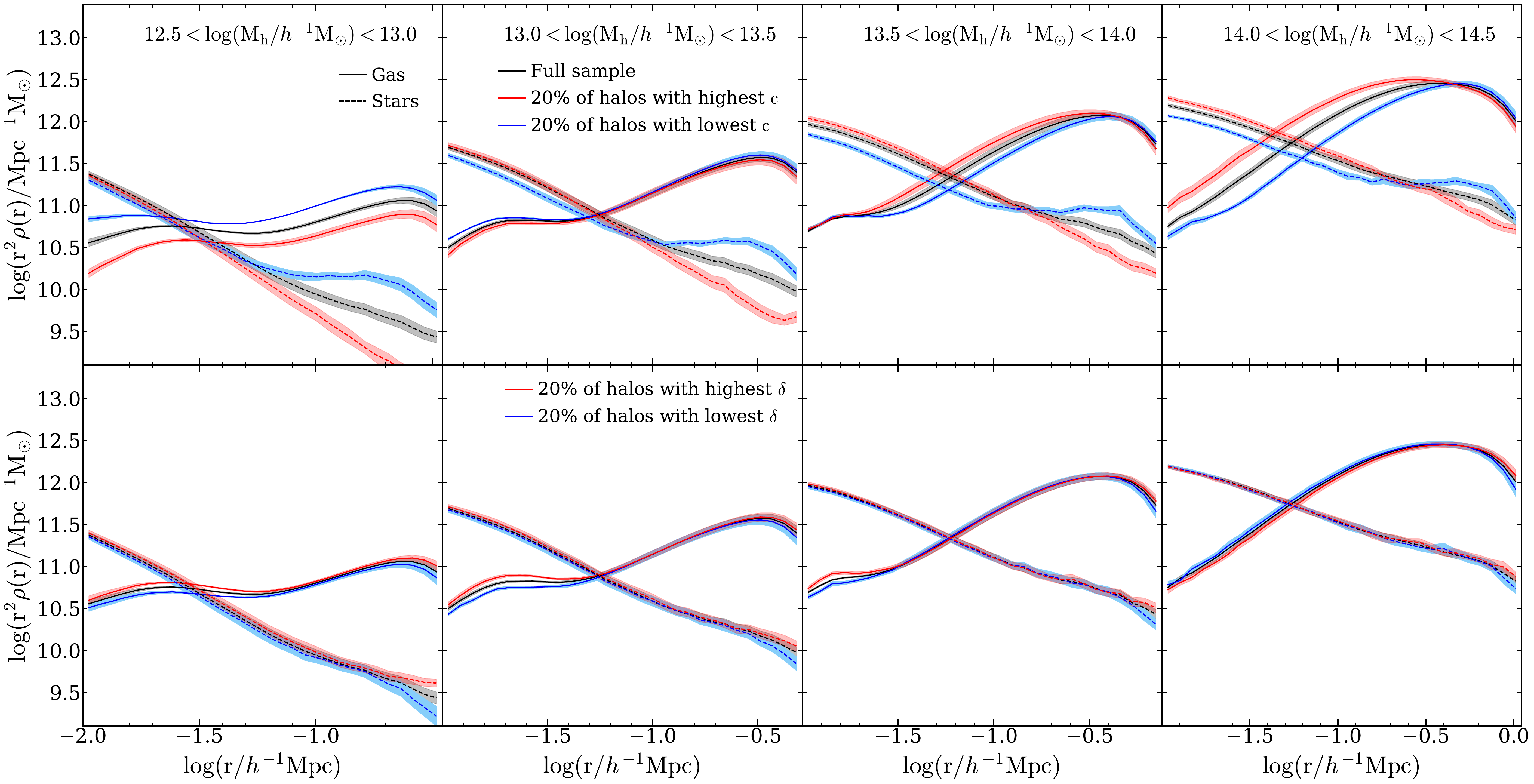}
    \caption{The secondary dependence of the density profile for the individual baryonic components at $\rm z = 0.0$. The top panels show the density profiles of stars (dashed lines) and gas (solid lines) for the full sample (black), and for the 20\% most (red) and least (blue) concentrated halos at fixed halo mass. Each column corresponds to a different halo mass interval. The bottom panels present the same measurements for the environmental dependence. Shaded regions indicate the statistical uncertainties on the mean value.}
    \label{fig:star_gas}
\end{figure*}

We begin by analyzing the density profiles of the individual baryonic components in halos with different secondary properties at fixed halo mass, which directly illustrate how these secondary dependencies manifest in the inner matter distribution of baryons. The hydrodynamical simulations track three types of baryons: gas, stars and black holes. Our analysis focuses on gas and stars, as the black hole mass contribution is negligible compared to the other components and has minimal influence on the total density profile. Figure~\ref{fig:star_gas} displays how the density profiles of gas and stars vary with secondary properties and halo mass. In the top panels, red and blue lines represent the top and bottom $20\%$ of halos ranked by concentration, while in the bottom panels they correspond to the most and least dense environments. Solid lines correspond to the gas profiles, and dashed lines indicate the contribution from stars. Consistent with our previous analysis, all signals are averaged from $10$ measurements in narrow halo mass bins of width $0.05$ dex, with shaded regions showing the error on the mean value (see section~\ref{subsec:measure}). The differences between the red/blue lines and the black lines (full sample) thus indicate the impact of the concentration and environment on the density profiles of baryons at fixed halo mass. 

As shown in the top panels of Fig.~\ref{fig:star_gas}, halo concentration modulates both the gas and stellar density profiles. For halos in the mass range of $12.5 < \log (\mathrm{M_h} / h^{-1}\mathrm{M_{\odot}}) <13.0 $, high-concentration halos exhibit significantly lower gas densities across all scales, whereas low-concentration halos contain a larger amount of gas. This difference directly contributes to the concentration dependence of the baryonic effects in the intermediate regions, where gas dominates the baryonic mass budget. With increasing halo mass, this trend reverses in the range $13.5 < \log (\mathrm{M_h} / h^{-1}\mathrm{M_{\odot}}) <14.0 $. The stellar components exhibit the opposite trend in the inner regions, where high-concentration halos contain slightly more massive central galaxies at fixed halo mass, consistent with the well-established concentration dependence of the stellar-to-halo mass relation (e.g., \cite{Matthee.2017}). This trend becomes more pronounced with increasing halo mass. At larger radial scales, however, more concentrated halos contain significantly fewer stars, with this trend weakening as halo mass increases. This effect is primarily driven by variations in the intracluster medium and satellite populations, consistent with ref.~\cite{Sergio.2019}. A comparison with Fig.~\ref{fig:c} suggests that the combination of the enhanced inner stellar profile and the reversal in gas density trends at high mass may contribute to the inversion of the total secondary concentration dependence signal in the same mass range. However, it is important to note that the effects on baryonic component profiles do not fully determine the total baryonic effect, as the dark matter back-reaction and the dependence of the DMO baseline profile on concentration must also be considered. 

As shown in the bottom panels of Fig.~\ref{fig:star_gas}, the influence of the environment is much weaker and exhibits less systematic trends compared to concentration. At the low-mass end, halos in denser environments exhibit slightly enhanced gas densities in their inner regions, an effect that diminishes at larger scales and vanishes for halos with $\log (\mathrm{M_h} / h^{-1}\mathrm{M_{\odot}}) > 14.0 $. The environmental dependence associated with the stars is small and comparable to the noise. Nevertheless, we see that halos in dense environments display slightly higher stellar densities across all scales in the lowest mass bin. These trends in the central region likely reflect the fact that, at fixed halo mass, halos in denser environments tend to host more massive central galaxies and more numerous satellites (e.g., \cite{Zehavi.2018, Artale.2018, Wang.2025}). Although this environmental imprint on baryonic components appears subtle, we remind the reader that it may lead to significant systematic effects in precision clustering analyses (see discussions in Section~\ref{subsec:env}).

\subsection{Impact of Baryonic Properties} \label{subsec:baryonic_property}

We further investigate the influence of different baryonic properties on the baryonic effects at fixed halo mass, focusing on three key quantities: stellar mass, black hole accretion rate, and gas mass. The stellar mass of the galaxy is a fundamental observable closely linked to baryonic processes, and correlates with both halo mass and concentration (e.g., \cite{Behroozi.2013, Matthee.2017, Kulier.2019}). As a direct tracer of cumulative baryonic condensation and integrated feedback history, it is expected to be tied to the baryonic modifications of halo inner density profiles. Therefore, we proceed to explore the impact of stellar mass on baryonic effects. In this work, we characterize this by the total stellar mass within the halo, rather than the often-used stellar mass of the central galaxy. This choice captures the contribution of the entire stellar content in the halos, including satellite populations and intracluster medium, which can influence the density profiles beyond the innermost regions. 

The black hole accretion rate, defined as the sum of instantaneous accretion rates of all black holes in a halo, regulates the mode of AGN feedback \cite{Begelman.2014}. High accretion rates typically trigger the thermal mode, characterized by efficient disk accretion and thermal energy injection into the surrounding gas. In contrast, low accretion rates are associated with the kinetic mode, which arises from radiatively inefficient disk accretion and drives energy injection in kinetic form, thereby more efficiently expelling gas from the halo (e.g., \cite{Blandford.1999, Yuan.2014}). We thus expect the black-hole accretion rate to be associated with variations in the baryonic modification of halo profiles. Finally, the gas mass, defined as the total mass of gas cells within twice the stellar half-mass radius, traces the gas content of the central galaxy and serves as an indicator of the fuel reservoir available for star formation. While we have previously established that the gas content is influenced by halo concentration and environment (see Fig.\ref{fig:star_gas}), we would like to determine whether this correlation is reflected in the total baryonic modification of halo density profiles.

\begin{figure*}
\centering
    \includegraphics[width=\textwidth]{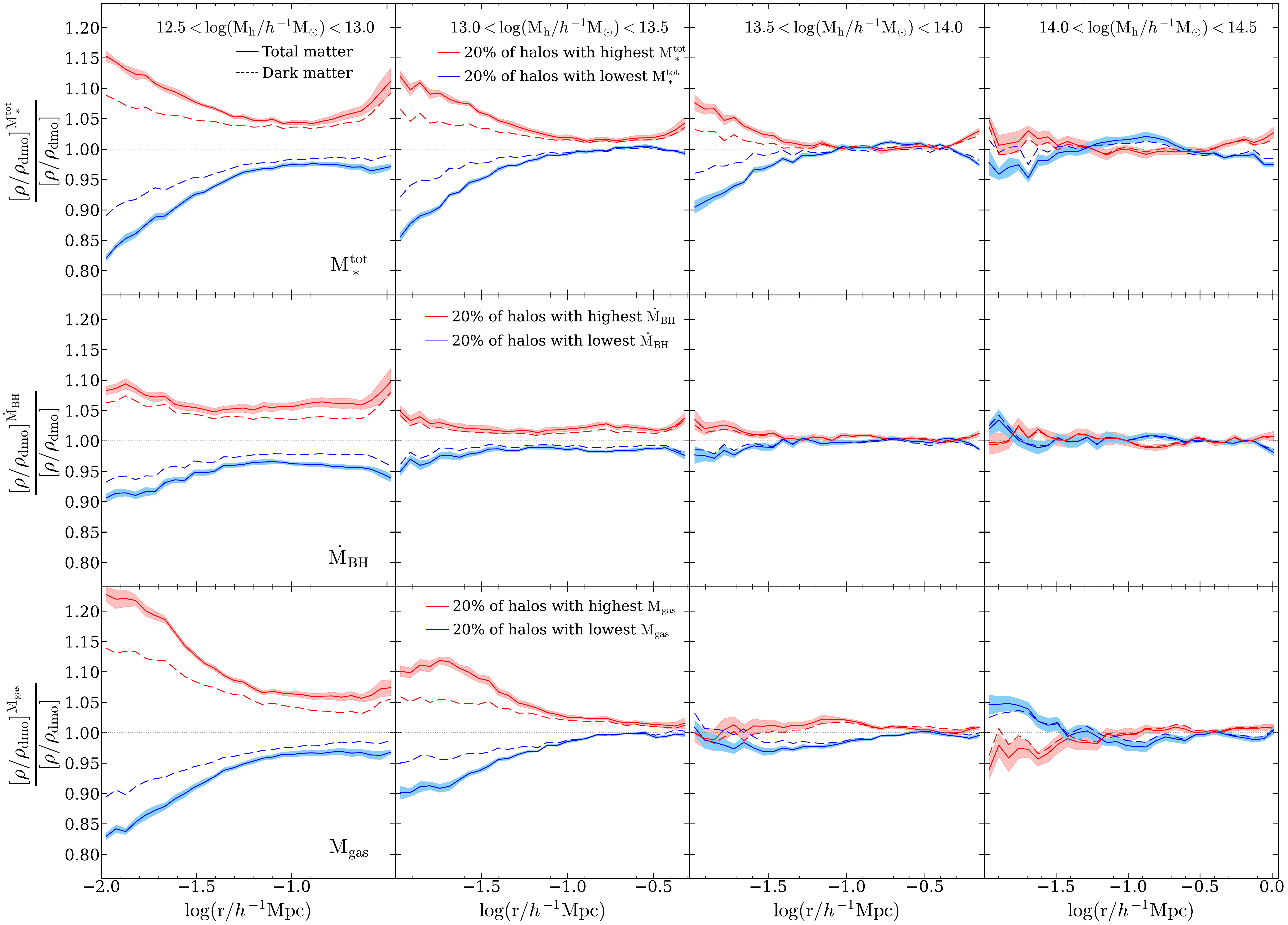}
    \caption{The dependence of baryonic effects on baryonic properties at $\rm z = 0.0$. Each column corresponds to a different halo mass bin. Halo subsets are selected by baryonic properties of the hydrodynamic halos for each matched pair. The solid lines represent the ratio of the average $\rho / \rho_{\mathrm{dmo}}$ for the total matter in the 20\% of halos with the highest (red) and lowest (blue) total stellar mass $\mathrm{M_*^{tot}}$ (top panels), black hole accretion rate $\mathrm{\dot{M}_{BH}}$ (middle panels), and gas mass $\mathrm{M_{gas}}$ (bottom panels), relative to the full halo population at fixed mass. The dashed lines correspond to the dark matter components. Shaded regions indicate the statistical uncertainties of the signals.}
    \label{fig:stellar_BH_gas}
\end{figure*}

We employ the same analysis framework described in Section~\ref{subsec:measure}, namely quantifying the baryonic effects via the ratio between the density profile of each hydrodynamical halo and that of its DMO counterpart, and comparing it across sub-samples of halos with different properties at fixed halo mass. However, the sub-samples are now defined based on the internal baryonic properties of the hydrodynamic halos. Figure~\ref{fig:stellar_BH_gas} presents the dependence of baryonic effects on these three baryonic properties, with each row corresponding to a different property. The red (blue) solid lines display the ratio of averaged $\rho/\rho_{\mathrm{dmo}}$ for the upper (lower) $20 \%$ of the property distribution relative to the full sample average at fixed halo mass at redshift $\rm z=0.0$. Dashed lines are the corresponding measurements for the dark matter components.

We first examine the influence of total stellar mass, $\rm M_*^{tot}$. As shown in the top panels of Fig.~\ref{fig:stellar_BH_gas}, halos with higher stellar mass display a significant enhancement in the ratio $\rho/\rho_{\mathrm{dmo}}$, reaching amplitudes of $\sim 15\%$ in the inner-most radii and decaying to $\sim 5\%$ on intermediate scales. This strong signal highlights the important role of stellar content in shaping baryonic effects beyond the primary dependence on halo mass. Halos with larger $\rm M_*^{tot}$ naturally contain higher baryonic density in the center, where stars are the dominant baryonic component (see Fig.~\ref{fig:star_gas}). In the intermediate region, the high stellar abundance leads to a weaker suppression of the density profile relative to the average, with these trends diminishing toward the high-mass end. The back-reaction signals, shown as dashed lines, exhibit similar trends, indicating that the dark matter density profile is similarly modulated by $\rm M_*^{tot}$. For completeness, we also examine the dependence on the stellar mass of central galaxies (not shown), finding consistent trends, albeit with slightly reduced amplitudes. The trends associated with stellar mass dependence differ significantly from the concentration dependence shown in Fig.~\ref{fig:c}. Although this is somewhat counterintuitive given the established relation between these properties (e.g., \cite{Montero.2016, Matthee.2017, Kulier.2019}), it can likely be understood through two effects. First, the correlation between concentration and stellar mass becomes weaker for the relatively massive halos considered in this work. Second, the total density profile is also shaped by the redistribution of dark matter and gas. This suggests that the observed concentration dependence reflects a more complex interplay of baryonic processes and cannot be simply explained by its correlation with stellar mass.

Next, we consider the dependence on the black hole mass accretion rate, $\rm \dot{M}_{\rm BH}$. As illustrated in the middle panels of Fig.~\ref{fig:stellar_BH_gas}, in the low-mass regime, halos with higher black hole mass accretion rates exhibit a pronounced enhancement of the ratio $\rho/\rho_{\mathrm{dmo}}$ within the central regions, reaching an amplitude of $\sim 8\%$, along with a sustained excess of $\sim 5\%$ in the intermediate regions. This behavior reflects a stronger central baryonic condensation and weaker suppression at intermediate scales, consistent with the expected impact of thermal-mode-dominated AGN feedback. Low-accretion halos, likely dominated by the kinetic mode, show the opposite trend with comparable magnitudes. These effects diminish with increasing halo mass, reflecting the reduced differential impact of AGN feedback in more massive systems.

The correlation between halo assembly history and black hole activity may help understand the observed secondary concentration dependence in Fig.~\ref{fig:c}. Highly concentrated halos typically form earlier and are dynamically more relaxed, which inherently leads to lower black hole accretion rates. Consequently, their AGN feedback is more likely to operate in the kinetic mode, which effectively removes gas from the central regions and produces stronger suppression in intermediate regions in lower-mass halos. This mechanism is consistent with the intermediate-scale trend in Fig.~\ref{fig:c}, where high-concentration halos exhibit stronger mass suppression. However, it does not fully account for the pronounced density enhancement in the inner regions, where the impact of concentration is much stronger than that of $\rm \dot{M}_{BH}$. For completeness, we also examine the dependence on the black hole mass, $\rm M_{BH}$, which can serve as a cumulative proxy for black-hole growth and integrated AGN activity and has been found to correlate more strongly with long-term galaxy quenching than instantaneous accretion indicators \citep{Piotrowska.2022}. We find qualitatively similar trends to those obtained for the black-hole accretion rate, consistent with the interpretation discussed above. Since this additional measurement does not introduce a qualitatively distinct trend, we do not show it separately. Taken together, these results suggest that the concentration dependence is partly associated with correlations among halo assembly history, current AGN feedback mode, and integrated AGN feedback history, with additional effects likely involved.

Finally, we investigate the influence of the mass of gas, $\rm M_{gas}$. As shown in the bottom panel of Fig.~\ref{fig:stellar_BH_gas}, the secondary dependence of baryonic effects on $\rm M_{gas}$ follows trends broadly similar to those seen for the stellar mass, yet with larger amplitudes at the low-mass end. Specifically, in the mass range $12.5< \log (\mathrm{M_h}/ h^{-1} \mathrm{M_{\odot}})< 13.0$, halos with higher gas mass exhibit a $\sim 20\%$ stronger enhancement of the central profile, while expelling less mass in the intermediate region with a weaker amplitude. As halo mass increases, the dependence on gas mass weakens rapidly and eventually reverses at the high-mass end. 

The significant gas mass dependence highlights the crucial role of gas in shaping the inner density profiles of low-mass halos. At fixed mass, a higher gas mass reflects a larger retained baryonic fraction, which naturally leads to stronger central mass contraction and indicates weaker feedback-driven mass expulsion at intermediate scales. This mechanism also provides further insight into the interpretation of the concentration dependence. More concentrated halos tend to have systematically lower gas content in the low-mass regime (see top panels of Fig.~\ref{fig:star_gas}), such that they undergo stronger profile suppression. Thus, this trend is also consistent with the pronounced concentration dependence observed in Fig.~\ref{fig:c}.

In summary, stellar mass, black hole accretion rate, and gas mass exhibit significant impact on baryonic effects at fixed halo mass, which provides valuable clues regarding the physical drivers of the concentration dependence.
Stellar mass and gas mass significantly modulate the amplitude of central enhancement, while the black hole accretion rate affects the level of intermediate-scale mass suppression through different AGN feedback modes. Despite these strong dependencies, many of the baryonic properties examined here are coupled and none of them reproduces the observed concentration dependence alone. Consequently, rather than acting strictly as isolated causes, these properties could be direct drivers of the profile variations, emerge as consequences of the same underlying physical processes, or a combination of both. And the pronounced concentration dependence cannot be explained solely by simple correlations with baryonic content. In general, these results suggest that the secondary dependence likely reflects the combined influence of multiple baryonic processes, linked to halo assembly history, that jointly shape the redistribution of matter within halos.

\section{Conclusion}
\label{sec:conclusion}

We have utilized the large-volume cosmological hydrodynamical simulation, MTNG, and its dark matter-only counterpart to investigate how secondary halo properties, such as the halo concentration and large-scale environment, modulate the baryonic effects on halo density profiles beyond halo mass. Specifically, we examine the ratio of the density profiles of matched halos in the hydrodynamical simulation and the corresponding DMO run. To isolate the secondary dependence, we compare the average baryonic effects of the top and bottom $20\%$ of halos, ranked by each property, with those of the full sample within narrow halo mass bins. We similarly analyze the back-reaction on the dark matter components. We further differentiate changes to the internal profile from the global mass shift, and independently examine the relationship between secondary signals and baryonic properties. Our main analysis focused on halo samples at redshift $\rm z = 0.0$, with an extension to $\rm z = 0.5$ to encompass the redshift range commonly probed in observations. We summarize our main results as follows:

\begin{itemize}

\item The impact of baryons enhances the halo density profile in the innermost regions while suppressing it at intermediate radii. The magnitude of this suppression decreases with increasing halo mass. The profile of the dark matter component exhibits a corresponding back-reaction effect, showing similar trends in the inner regions, but with a significantly weaker impact at intermediate radii.

\item Halo concentration induces a pronounced impact on the baryonic effects at fixed halo mass. More concentrated halos exhibit weaker central enhancement and stronger intermediate suppression of their profile due to baryonic effects, whereas the trend reverses at the high-mass end. The effect is strongest in the lowest-mass halos, reaching $\sim 15\%$ in the innermost regions and $\sim 8\%$ at intermediate radii for the $\rm z = 0.0$ samples, with the overall amplitude decreasing for higher halo mass. 

\item The large-scale environment exhibits a weaker impact on the baryonic effects at fixed halo mass. This environmental dependence is nearly scale-independent, with a slight increase in the density profile of halos in denser regions and a slight decrease for halos in under-dense regions, both at the $\sim 2\%$ level. These effects diminish for increased halo mass and also show no correlation with halo concentration. 

\item Baryonic effects also depend on additional secondary properties, such as halo spin and velocity dispersion, both exhibiting significant effects (see Appendix~\ref{app:spin}). Halos with higher spin show a stronger inner enhancement and a weaker intermediate suppression at lower halo mass, whereas the trend weakens and reverses with increasing halo mass. The velocity dispersion displays trends qualitatively similar to those of concentration, albeit with slightly reduced amplitudes.

\item Halos at $\rm z= 0.5$ exhibit qualitatively similar baryonic effects and secondary dependencies compared to those at $\rm z = 0.0$. The environmental dependence weakens slightly, whereas the concentration dependence of the intermediate suppression becomes more pronounced for lower-mass halos.

\item We find that the secondary concentration dependence can be characterized as the combination of: (1) a predominantly mass-independent restructuring of the density profile, and (2) a scale-independent overall mass shift that varies with halo mass. In contrast, the secondary environmental dependence is only dominated by a small mass-dependent shift in the total halo mass. 

\item The secondary dependencies are reflected in the distribution of baryonic components within halos. In low-mass halos, higher concentration is associated with reduced gas and stellar content. This trend weakens and reverses with increased halo mass. The environmental dependence is weak and shows no clear systematic trend, although halos in denser regions tend to contain slightly more gas.

\item The baryonic effects on halo density profiles show significant dependence on internal baryonic properties, such as stellar mass, black hole accretion rate, and gas mass. While these properties modulate the shape and amplitude of the signal, they do not directly account for the observed concentration dependence, suggesting that the latter arises from a more complex interplay of baryonic processes linked to halo assembly history.

\end{itemize}

Compared with previous studies on baryonic modifications to halo density profiles (e.g., \cite{Cui.2012, Cui.2014, Schneider&Teyssier.2015, G.Arico.2020, G.Arico.2021}) and the influence of secondary properties (e.g., \cite{Beltz-Mohrmann.2021, Y.Wang.2024, Elbers.2025}), our work provides a more systematic characterization of the secondary dependence of baryonic effects on halo density profiles at fixed halo mass. Leveraging the large volume and high resolution of the MTNG simulation, we achieve improved statistical precision and stricter control over the halo mass, while exploring a wider range of secondary properties and assessing their correlation with different baryonic components. We therefore view our results as valuable guidance for future extensions of baryonic correction models. In particular, our findings suggest that halo concentration may play an important role in modeling baryonic effects beyond mass-only prescriptions and could be incorporated through concentration-dependent profile parameters controlling the central enhancement, matter redistribution, and dark-matter response. Although the environmental dependence appears weaker, it may still be relevant for precision applications because percent-level environment-dependent residuals could bias clustering or weak-lensing predictions, and this dependence could be modeled through an environment-dependent mass correction. While developing such a calibrated prescription is beyond the scope of this work, it represents a natural direction for future modeling.

We note that our quantitative results are specific to the MillenniumTNG, as the level of baryonic effects varies with the galaxy formation model. Previous comparisons across hydrodynamical simulations have shown that different feedback prescriptions can produce substantial variations in the displacement of gas within and beyond the virial radius, the back-reaction on the dark matter, and the resulting modification of matter clustering, particularly arising from efficient AGN-driven outflows and the stellar feedback (e.g., \cite{Ayromlou.2023, Gebhardt.2024, Gebhardt.2026}). Moreover, baryonic effects may couple to cosmology through changes in halo baryon fractions, concentrations, and binding energies (e.g., \cite{Schaller.2025,Pranjal.2025, Elbers.2025}). While these model variations may alter the strength and characteristic scales of the secondary dependencies reported here, the qualitative trends are expected to hold more generally, given that our analysis focuses on relative differences between halo populations at fixed mass. In future work, we plan to extend this study to additional simulation suites with different galaxy formation prescriptions and cosmologies. Ultimately, we aim to incorporate these secondary dependencies into more comprehensive baryonic modeling frameworks, to enable more accurate cosmological inferences from next-generation large-scale structure observations.

\vspace{4cm}

\section*{Data Availability}

The MillenniumTNG simulations are expected to be publicly available later this year and accessible at \url{www.mtng-project.org}. Data directly related to this publication is available upon reasonable request.

\vspace{4cm}

\acknowledgments

We thank Volker Springel and the other members of the MillenniumTNG team for providing proprietary data access. We thank Kai Wang for insightful discussions and helpful comments. We also appreciate the constructive comments and thoughtful suggestions provided by the anonymous referee, which greatly improved the presentation of the paper. S.C. acknowledges the support of a ``Ram\'on y Cajal'' fellowship (RYC2023-043783-I). S.B. is supported by the UKRI Future Leaders Fellowship (MR/V023381/1 and UKRI2044).

\vspace{8cm}

\appendix

\section{Scatter of the measurement on individual halos} \label{app:scatter}

\begin{figure*}
\centering
    \includegraphics[width=\textwidth]{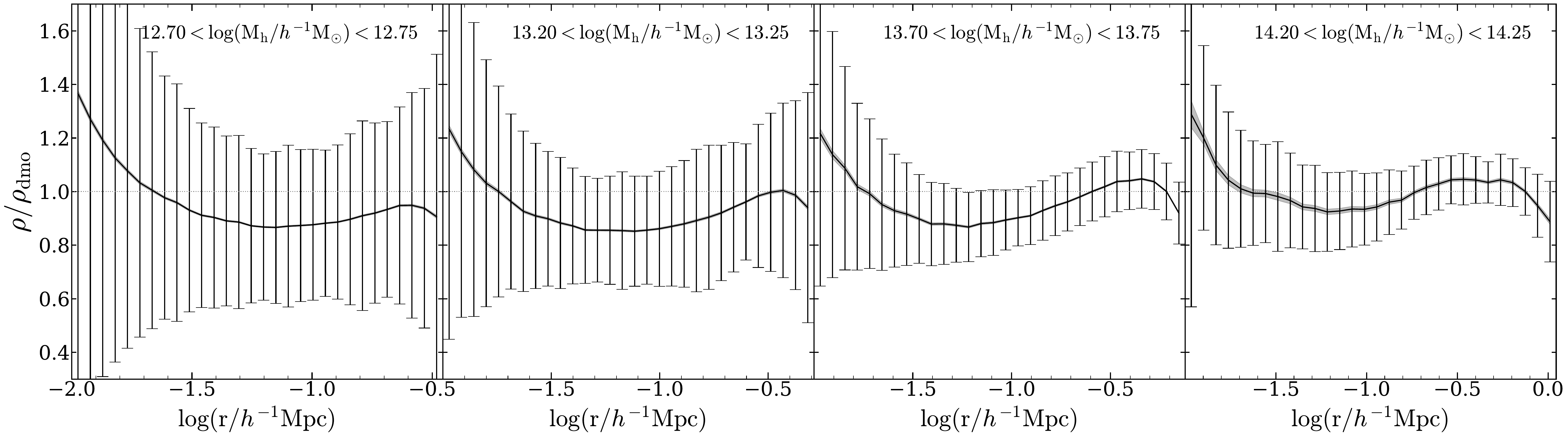}
    \caption{The scatter of baryonic effects on the density profiles of halos in representative $0.05$ dex mass bins. Black lines show the mean value of $\rho/\rho_{\mathrm{dmo}}$ for each bin, with the corresponding mass range labeled in each sub-panel. Error bars represent the halo-to-halo scatter (standard deviation) within each halo sample, while the shaded regions show the corresponding statistical uncertainties on the mean.}
    \label{fig:scatter}
\end{figure*}

As described in Section~\ref{subsec:measure}, measurements within each narrow $0.05$ dex mass bin are derived by averaging the $\rho/\rho_{\mathrm{dmo}}$ ratios of individual halos in the selected sub-samples. The final signal for a $0.5$ dex mass interval is then computed by averaging the results from $10$ such consecutive bins. While the intrinsic scatter across individual halos within a bin of $0.05$ dex width can be considerable, the large sample size ensures that the statistical uncertainty on the mean remains small. For completeness, we show in Figure~\ref{fig:scatter} the baryonic effects for a representative $0.05$ dex bin from each mass interval together with halo-to-halo scatter (error bars) in $\rho/\rho_{\mathrm{dmo}}$, compared to the statistical uncertainty on the mean (shaded region). Low-mass halos exhibit larger scatter, likely reflecting the greater intrinsic variation in their secondary properties at fixed halo mass compared to more massive halos. As is evident, however, the resulting standard error on the mean remains significantly small across all bins and becomes negligible at the low-mass end.

This demonstrates a key advantage of the MTNG simulation: its large volume and high resolution provide a sufficiently high halo count to enable statistically robust measurements even within narrow mass bins. We initially did a similar analysis using Illustris-TNG300 and its corresponding DMO run. The smaller halo sample size in TNG300-1 resulted in substantially larger statistical uncertainties using the same narrow-binning scheme, limiting its use for robust measurements of these subtle secondary dependencies. Finally, we emphasize that all analyses in this work are based on the statistical average derived from sufficiently large samples. While baryonic effects on the density profile physically originate at the level of individual halos, the secondary dependence signals reported here reflect population-level trends and should not be interpreted as deterministic predictions for individual systems.

\section{Secondary dependence on halo spin and velocity dispersion } \label{app:spin}

We present complementary results for two additional secondary properties: halo spin and halo velocity dispersion. Halo spin is an important element in galaxy formation models, as it traces the angular momentum of dark matter halos and is linked to galaxy formation and morphology (e.g., \cite{Bullock.2001,Maller.2002, Bett.2007}. Various studies of assembly bias also identify halo spin as a secondary halo property that correlates with halo assembly history (e.g., \cite{Gao.2007, Faltenbacher.2010, Lacerna.2012}). Given that variations in angular momentum and merger history may influence the distribution of baryons and the efficiency of feedback processes within halos, we extend our analysis to explore the dependence of baryonic effects on halo spin. Specifically, we adopt the dimensionless spin parameter defined by Ref.~\cite{Bullock.2001}: $\rm \lambda = J/(\sqrt{2} M_{200c}V_{200c}R_{200c})$, where $\rm J$ is the angular momentum within a sphere of virial radius $\rm R_{200c}$, $\rm M_{200c}$ is the corresponding virial mass and $\rm V_{200c}$ is the halo circular velocity at that radius. 

The velocity dispersion of the halo, $\sigma$, is defined as the one-dimensional dispersion of the velocities of all dark matter particles bound to the DMO halo. It characterizes the internal kinematics and traces the depth of the gravitational potential of the halo, thereby serving as a proxy for cluster mass and connecting to observable baryonic signatures (e.g., \cite{Biviano.2006,Evrard.2008, Munari.2013}). In addition, velocity dispersion has been shown to correlate with halo age and concentration at fixed halo mass, thereby leading to variations in halo clustering beyond mass (e.g., \cite{Gao.2007, Faltenbacher.2010, Montero.2025}). Motivated by these connections, we include velocity dispersion as an additional secondary property in our analysis.

\begin{figure*}
\centering
    \includegraphics[width=\textwidth]{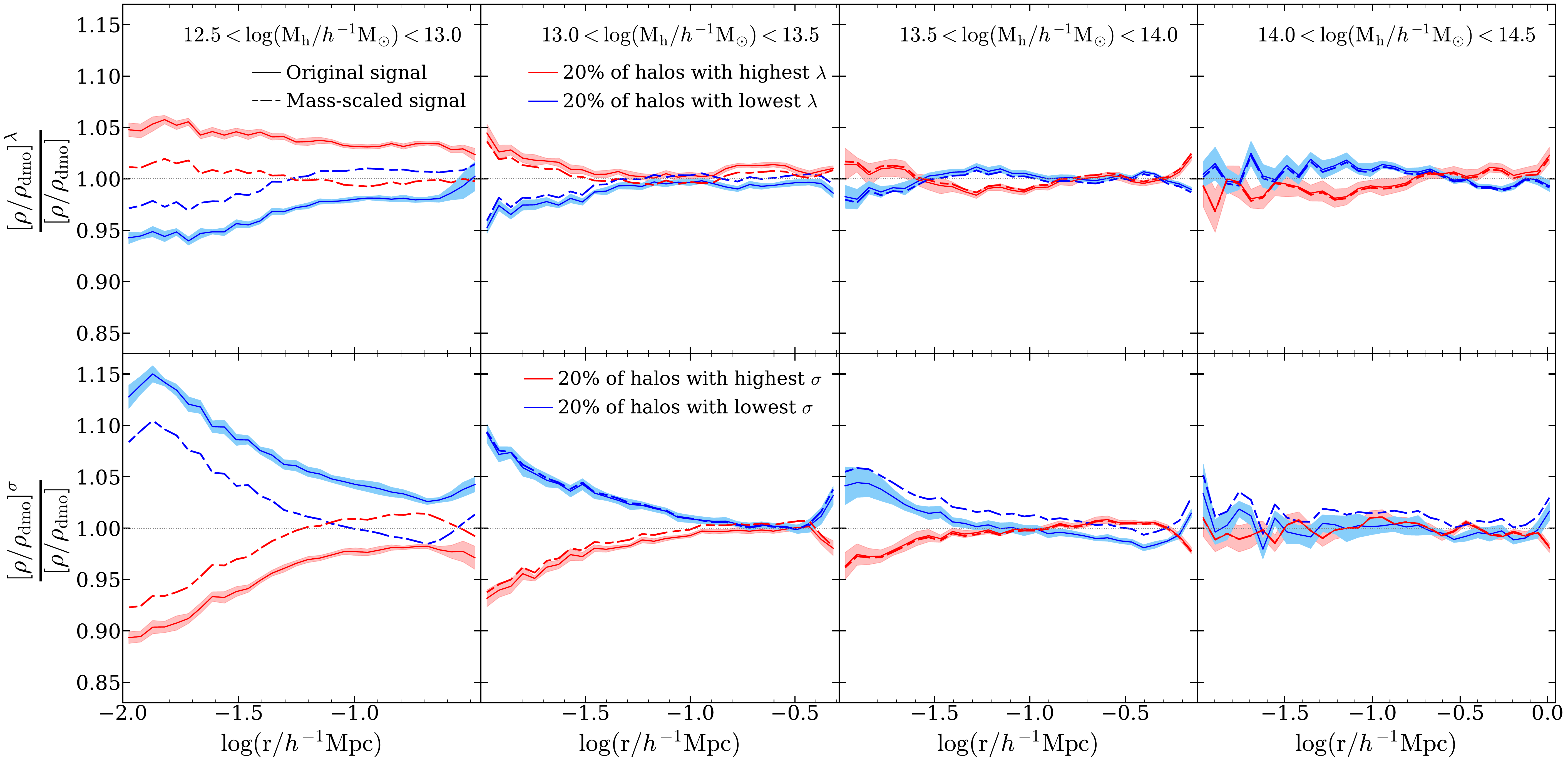}
    \caption{The secondary dependence of baryonic effects on halo spin (top panels) and velocity dispersion (bottom panels) at $\rm z = 0$. This figure is analogous to the Fig.~\ref{fig:norm}, but for halo sub-samples selected using halo spin and velocity dispersion. The solid lines and dashed lines represent the original secondary dependence and the mass-scaled signals. Shaded regions indicate the statistical uncertainties on the mean value of the original signals.}
    \label{fig:spin_vel}
\end{figure*}

Following the same methodology as described in Section~\ref{subsec:measure}, we rank matched halo pairs by their spin or velocity dispersion measured in the DMO simulation within $0.05$ dex halo mass bins, and compare the baryonic effects of the top and bottom $20\%$ sub-samples to those of the full sample. The averaged result across ten adjacent bins is taken as the representative signal for each of the four $0.5$ dex halo mass intervals. Figure~\ref{fig:spin_vel} shows the dependence of baryonic effects on halo spin (top panels) and velocity dispersion (bottom panels) at $z = 0.0$. The solid lines represent the ratio between the $\rho /\rho_{\mathrm{dmo}}$ of the extreme sub-samples, relative to the full population at fixed halo mass. We also examine the signals on the mass-scaled profiles, shown as dashed lines, which remove the overall halo mass change from the total baryonic effects following the procedure of Section~\ref{subsec:normed}. The secondary dependencies of back-reaction effects mirror the trends shown in the total matter profile and are therefore omitted for visual clarity. The shaded regions indicate the statistical uncertainties of the original signals, while the uncertainties of the mass-scaled signals (not shown) are of comparable magnitude.

Examining the halo spin dependence, shown in the top panels of Fig.~\ref{fig:spin_vel}, a clear signal is apparent in the lowest mass bin ($12.5 < \log (\mathrm{M_h} / h^{-1}\mathrm{M_{\odot}}) <13.0 $), with the amplitude decreasing with scale. High spin halos exhibit stronger matter condensation ($\sim 5 \%$ enhancement) in the central region and a reduced mass suppression by about $2.5\% - 5\%$ at intermediate to large scales. This trend weakens with increasing mass and reverses at the high mass end. After scaling the halo profiles by $\rm M_{h}/ M_{h, dmo}$, the signal for the lowest-mass halos preserves the original trend in the inner regions but reverses at intermediate scales. This suggests that high-spin halos not only have a higher overall halo mass relative to their DMO counterparts but also undergo stronger baryonic processes that lead to a pronounced redistribution of baryons toward the center. As halo mass increases, the difference between the solid and dashed lines diminishes, indicating that the impact of halo spin on the total mass shift becomes negligible in massive systems.

As shown in the bottom panels of Fig.~\ref{fig:spin_vel}, the dependence on velocity dispersion resembles that of halo concentration (see bottom panels of Fig.~\ref{fig:c}), albeit with a slightly reduced amplitude in the lowest mass bin. This is consistent with the concentration-velocity dispersion relation, with more concentrated halos having larger velocity dispersions (e.g., \cite{Faltenbacher.2007, Ragone.2010}). Specifically, baryonic effects on halos with lower velocity dispersion tend to enhance central mass accretion and weaken mass expulsion in the intermediate regions. Comparing the mass-scaled and the original signals, the velocity dispersion dependence of total mass shift also follows similar trends as halo concentration.

To further assess whether the trends associated with spin and velocity dispersion correlate with concentration, we performed the same analysis at fixed halo mass and concentration (not shown), following a similar approach to that adopted for environment in Fig.~\ref{fig:double_env}. We find that the spin dependence persists with only a minor reduction in amplitude, while the residual dependence on velocity dispersion largely vanishes. This suggests that halo spin may act as an additional parameter for extending the baryonification model, whereas the impact of velocity dispersion can be effectively captured via halo concentration alone. We also extended the measurements to $\rm z=0.5$, finding trends qualitatively consistent with the $z=0.0$ results. As the observed redshift evolution closely follows the trends established in Fig.~\ref{fig:all_0.5} and discussed in Section~\ref{subsec:higher_z}, we do not duplicate those details here.

\vspace{1cm}

\bibliographystyle{JHEP}
\bibliography{Baryonic.bib}

\end{document}